\documentclass[12pt]{article}
\pdfoutput=1

\usepackage[utf8]{inputenc}
\usepackage{graphics,amssymb,float}
\usepackage{graphicx}
\usepackage{color}
\usepackage{amsmath}
\usepackage{xspace}
\usepackage{placeins}
\usepackage{dcolumn}
\usepackage{bm}

\usepackage{cite}
\usepackage{hyperref}
\usepackage{indentfirst}  
\usepackage{rotating} 
\usepackage{longtable,multirow}

\newcommand{\MET}{E_T^{\rm miss}}

\def\lsim{\mathrel{\raise.3ex\hbox{$<$\kern-.75em\lower1ex\hbox{$\sim$}}}}
\def\gsim{\mathrel{\raise.3ex\hbox{$>$\kern-.75em\lower1ex\hbox{$\sim$}}}}
\def\ifmath#1{\relax\ifmmode #1\else $#1$\fi}

\def\smodels{{\sc SModelS}}
\def\smodelsnn{{\sc SModelS}}
\def\smodelsv{{\sc SModelS}\,v1.1.1}
\def\fastlim{{\sc Fastlim}}
\def\eg{{\it e.g.}}
\def\ie{{\it i.e.}}

\newcommand{\met}{\ensuremath{E_{T}^{\text{miss}}}}
\newcommand\ZeroLepton{{0-lepton + 2--6 jets + $\met$}}
\newcommand\Multijet{{0-lepton + 7--10 jets + $\met$}}
\newcommand\OneLeptonStrong{{1-lepton + jets + $\met$}}
\newcommand\SSThreeLepton{{SS/3-leptons + jets + $\met$}}
\newcommand\TauStrong{{$\tau(\tau/\ell)$ + jets + $\met$}}
\newcommand\ThreeBjets{{0/1-lepton + 3\emph{b}-jets + $\met$}}
\newcommand\Monojet{{Monojet}}
\newcommand\ZeroLeptonStop{{0-lepton stop}}
\newcommand\OneLeptonStop{{1-lepton stop}}
\newcommand\TwoLeptonStop{{2-leptons stop}}
\newcommand\StopMonojet{{Monojet stop}}
\newcommand\StopZ{{Stop with $Z$ boson}}
\newcommand\TwoBjet{{2\emph{b}-jets + $\met$}}
\newcommand\TBmet{{$tb$+$\met$, stop}}
\newcommand\TwoLepton{{2-leptons}}
\newcommand\ThreeLepton{{3-leptons}}
\newcommand\FourLepton{{4-leptons}}
\newcommand\TwoTau{{2-$\tau$}}
\newcommand\OneLeptonHiggs{{$\ell h$}}
\newcommand\DisappearingTrack{{Disappearing Track}}
\newcommand\LLSparticles{{Long-lived particle}}
\newcommand\HiggsToTauTau{{$H/A\rightarrow\tau^+\tau^-$}}

\begin{document}
\begin{center}

\vspace*{-1.4cm}
\begin{flushright}
HEPHY-PUB 990/17
\end{flushright}

\vspace*{2cm}

{\Large\bf On the coverage of the pMSSM \\[2mm] by simplified model results} 

\vspace*{1cm}

\renewcommand{\thefootnote}{\fnsymbol{footnote}}

{\large 
Federico~Ambrogi$^{1}$\footnote[1]{Email: federico.ambrogi@oeaw.ac.at},
Sabine~Kraml$^{2}$\footnote[2]{Email: sabine.kraml@lpsc.in2p3.fr},
Suchita~Kulkarni$^{1}$\footnote[3]{Email: suchita.kulkarni@oeaw.ac.at}, \\[1mm]
Ursula~Laa$^{2,3}$\footnote[4]{Email: ursula.laa@lpsc.in2p3.fr},
Andre~Lessa$^{4}$\footnote[5]{Email: andre.lessa@ufabc.edu.br}, 
Wolfgang~Waltenberger$^{1}$\footnote[6]{Email: wolfgang.waltenberger@oeaw.ac.at}
} 

\renewcommand{\thefootnote}{\arabic{footnote}}

\vspace*{1cm} 

{\normalsize \it 
$^1\,$Institut f\"ur Hochenergiephysik,  \"Osterreichische Akademie der Wissenschaften,\\ Nikolsdorfer Gasse 18, 1050 Wien, Austria\\[1mm]
$^2\,$Laboratoire de Physique Subatomique et de Cosmologie, Universit\'e Grenoble-Alpes,
CNRS/IN2P3, 53 Avenue des Martyrs, F-38026 Grenoble, France\\[2mm]
$^3\,$LAPTh, Universit\'e Savoie Mont Blanc, CNRS, B.P.110 Annecy-le-Vieux,\\ F-74941 Annecy Cedex, France\\[2mm]
$^4\,$Centro de Ci\^encias Naturais e Humanas, Universidade Federal do ABC,\\ Santo Andr\'e, 09210-580 SP, Brazil\\[2mm]
}

\vspace{6mm}

\begin{abstract}
We investigate to which extent the SUSY search results published by ATLAS and CMS in the context of simplified models 
actually cover the more realistic scenarios of a full model. Concretely, we work within the phenomenological MSSM (pMSSM) 
with 19 free parameters and compare the constraints obtained from \smodelsv\ with those from the ATLAS pMSSM study 
in arXiv:1508.06608. 
We find that about 40--45\% of the points excluded by ATLAS escape the currently available simplified model constraints. 
For these points we identify the most relevant topologies which are not tested by the current simplified model results. 
In particular, we find that topologies with asymmetric branches, including 3-jet signatures from gluino-squark associated 
production, could be important for improving the current constraining power of simplified models results.
Furthermore, for a better coverage of light stops and sbottoms, constraints for decays via heavier neutralinos and charginos, 
which subsequently decay visibly to the lightest neutralino are also needed.
\end{abstract}

\end{center}


\clearpage

\section{Introduction}\label{sec:intro}

Simplified models~\cite{hep-ph/0703088, 0810.3921, 1105.2838, Okawa:2011xg, 1301.2175} have become one of the standard methods 
to interpret searches for physics beyond the Standard Model (BSM).  They reduce full models with dozens of particles and a plethora of 
parameters to subsets with just a handful of new states. 
The virtue of simplified model spectra (SMS), namely that a full model decomposes into many different SMS, also defines their main challenge: 
depending on the complexity of the mass and decay patterns, a full model may not be fully reconstructed by SMS. 
The question that arises is to what extent full models can indeed be constrained by SMS results. 

In this article, we address this question for a 19-parameter version of the minimal supersymmetric standard model, the so-called phenomenological MSSM~\cite{Djouadi:1998di}, or pMSSM for short. Our work is based on the  ATLAS pMSSM study~\cite{Aad:2015baa}, in which the points from an extensive pMSSM scan were tested against the constraints from 22 ATLAS searches from LHC Run~1. 
ATLAS made the SLHA spectra of the whole scan publicly available on HepDATA~\cite{ATLASpMSSMhepdata} together with information about which point is excluded by which analyses. This is extremely useful information, which we here use to test the constraining power of SMS results by means of \smodels~\cite{Kraml:2013mwa,Ambrogi:2017neo}. 

\smodels\ is an automatised tool for interpreting simplified model results from the LHC.
It decomposes collider signatures of new physics featuring a $\mathbb{Z}_2$-like symmetry into simplified model topologies, 
using a generic procedure where each SMS is defined by the vertex structure and the Standard Model (SM) final state particles; 
BSM particles are described only by their masses, production cross sections and branching ratios. The weights of the various topologies, 
computed as production cross section times branching ratios,  
are then compared against a large database of experimental constraints.  
This procedure takes advantage of the large number of simplified models already constrained by official ATLAS and CMS
results and does not require Monte Carlo event simulation, thus providing a fast way of confronting a full BSM model
with the LHC constraints.
Furthermore, ``missing'' topologies, which are not covered by any of the experimental constraints, are also identified
and provided as an output of \smodels.

The tool can be used for testing any BSM scenario with a $\mathbb{Z}_2$-like symmetry as long as all heavier 
odd particles (cascade-)decay promptly to the lightest one, which should be electrically and colour neutral.%
\footnote{The treatment of charged tracks is also possible in the context of simplified models~\cite{Heisig:2015yla} and will be available in future versions of \smodels.} 
It has been applied to a number of minimal and non-minimal supersymmetric (SUSY) models 
in~\cite{Kraml:2013mwa,Belanger:2013pna,Barducci:2015zna,Arina:2015uea} 
but may also be used for non-SUSY models, see e.g.~\cite{Edelhauser:2015ksa,Kraml:2016eti}.
The underlying assumption~\cite{Kraml:2013mwa} that differences in the event kinematics 
(\eg\ from different production mechanisms or from the spin of the BSM particle)
do not significantly affect the signal selection efficiencies has also been investigated. 
For example, the effects of alternative production channels in squark simplified models were studied in~\cite{Edelhauser:2014ena}. 
The effect of a different spin structure was studied for the case of the dijet+MET final state in~\cite{Edelhauser:2015ksa}, 
for the dilepton+MET final state in~\cite{Arina:2015uea} and for $t\bar{t}$+MET final states in~\cite{Kraml:2016eti}. 
A comprehensive study of how well a full model like the MSSM is actually covered by SMS constraints is, however, still missing. 
This gap we want to fill with the present paper. 

We first describe the setup of the analysis in Section~\ref{sec:setup}. Our results are presented in Section~\ref{sec:results}, where we discuss 
the exclusion obtained with \smodelsv\ as compared to ATLAS and how it is improved when including efficiency maps in addition to upper limit maps. 
Moreover, we discuss why a certain part of the parameter space, despite being excluded by the ATLAS study, is not excluded by 
(the currently available) SMS results. In particular, we analyse the importance of asymmetric decay branches and long cascade decays  
to understand the potential for increasing the coverage, and we point out a number of important SMS beyond those typically considered by the experimental collaborations. Conclusions are presented in Section~\ref{sec:conclusions}. 
Appendices A and B contain useful additional material on the missing topologies discussed in the paper.

\section{Setup of the analysis}\label{sec:setup}

In~\cite{Aad:2015baa} ATLAS has analysed in total more than 310k pMSSM parameter points with SUSY masses below 4 TeV and a neutralino as the lightest SUSY particle (LSP). These points from an extensive scan, based on previous phenomenological studies \cite{Berger:2008cq,CahillRowley:2012cb,CahillRowley:2012kx,Cahill-Rowley:2014twa}, 
satisfy constraints from previous collider searches, flavor and electroweak (EW) precision measurements, cold dark matter relic density and direct dark matter searches.  
In addition, the mass of the light Higgs boson was required to be between $124$ and $128$~GeV.
These points were classified into three sets according to the nature of the LSP: bino-like (103410 points), wino-like (80233 points) and higgsino-like (126684 points).  
About 40\% of all these points were excluded by at least one of the 22 ATLAS Run~1 searches. 

The points excluded by ATLAS are the center of interest of our study: 
our aim is to compare the exclusion coverage obtained using SMS results only to  that from full event simulation. (In the following we mean by ``coverage'' the fraction of points excluded by ATLAS which is also excluded by \smodels.)
We restrict our analysis to the sets with bino-like or higgsino-like LSP, neglecting points with a wino-like LSP, as most of them lead to a displaced vertex signature, which cannot be studied with the current version of \smodels. 
We further remove points from the bino- and higgsino-like LSP data sets if they contain any long lived sparticles---this concerns however only a small number of points. 
Likewise, points which ATLAS found to be excluded only by heavy Higgs searches are also not considered here, as such searches are not 
treated in \smodels\ for the time being.
This selection leaves us with 38575 parameter points with a bino-like LSP and 45594 parameter points with a higgsino-like LSP to be tested with \smodels. 

We use the latest version of \smodels, v1.1.1, which works with upper limit (UL) and efficiency map (EM) type results, see~\cite{Ambrogi:2017neo}. 
The cross sections for all points are calculated with the \smodels\ cross section calculator interfaced to Pythia\,8.2~\cite{Sjostrand:2006za,Sjostrand:2014zea} and NLLfast~\cite{nllfast,Beenakker:1996ch,Kulesza:2008jb,Kulesza:2009kq,Beenakker:2009ha,Beenakker:2011fu,Beenakker:1997ut,Beenakker:2010nq}. (The exception are the cross sections for slepton-pair production, for which we use Pythia\,6.4~\cite{Sjostrand:2006za} because they are not computed correctly in Pythia\,8.226.) 
Electroweak cross sections are thus computed at leading order while strong productions are computed at NLO+NLL order.
Given the information on cross sections ($\sigma$) and decay branching ratios (BR) in the SLHA~\cite{Skands:2003cj} files, \smodels\ computes $\sigma\times {\rm BR}$ for each topology that occurs. Topologies are characterised by the SM particles originating from each vertex, and the mass vector of the SUSY particles in the decays. In order to avoid dealing with a large number of irrelevant processes, \ie\ to save CPU time, topologies for which 
$\sigma\times {\rm BR}<{\tt sigmacut}$, with ${\tt sigmacut} = 0.03$~fb, are discarded. 

In addition, if the mass gap between mother and daughter particles is small, the decay products will be too soft to be detected at the LHC. This is taken care of by the so-called ``mass compression'' in \smodels, discarding any SM particle coming from a vertex for which the mass splitting of the R-odd particles is less than a certain threshold. We use the default value of 5~GeV as the minimum required mass difference for the decay products to be visible.

After the decomposition, the weights  (\ie\ $\sigma\times {\rm BR}$) of the SMS components of each point are 
rescaled by the corresponding efficiencies (see~\cite{Ambrogi:2017neo} for more details) and matched with  
the experimental results in the database. In case of UL maps, this is a direct comparison of  individual weights and the cross section upper limit
for a given simplified model component or topology.
In case of EMs, the weights of several topologies can be combined and may contribute to a specific signal region of a given analysis;
it is then the combined signal cross section for the most sensitive signal region (\ie\ the signal region with the best expected limit)
which is compared against the experimental limit.  Hence using efficiency maps can significantly improve
the constraining power of simplified models.
See the \smodelsv\ manual~\cite{Ambrogi:2017neo} for a detailed explanation of the procedure. 

For a fair comparison with~\cite{Aad:2015baa}, we employ only the 8~TeV results in the v1.1.1 database.  
In order to maximise the coverage by SMS, we consider however also CMS 8~TeV results, as they may give complementary constraints. 
This is justified because ATLAS and CMS SUSY searches largely consider the same final states and have very similar reach. 
We also note that the official ATLAS and CMS Run~1 results available in \smodels\ were augmented with several `home-grown' EMs in the v1.1.1 database to increase the coverage, and we further extend this database with \fastlim-1.0~\cite{Papucci:2014rja} EMs as explained in~\cite{Ambrogi:2017neo}.  The complete list of analyses and results included in the v1.1.1 database can be consulted at~\cite{smodels:listofanalyses}.

A comparison of the analyses considered in \cite{Aad:2015baa} and the SMS results included in \smodels\ v1.1.1 is given in 
Table~\ref{tab:AtlasSearchesInPaper}. The analyses covered by the \fastlim\ EMs are listed in Table~\ref{tab:FastlimAnalyses}. 
Here note that in \smodelsv\ efficiencies with a relative statistical uncertainty greater than 25\% 
are set to zero and, moreover, zero-only EMs are discarded per default. Therefore, from the 264 EMs of \fastlim-1.0,   
which are based on 11 ATLAS conference notes, used in practice are 163 EMs from 9 conference notes. 
The CMS analyses included in the v1.1.1 database are listed in Table~\ref{tab:CMSAnalyses}.

\smodels\ reports its results in the form of $r$-values, defined as the ratio of the theory prediction over the observed 95\% confidence level (CL) upper limit, for each experimental constraint that is matched in the database.  We consider as excluded all points for which at least one $r$-value equals or exceeds unity ($r_{\rm max} \ge 1$).\footnote{We note that for staying strictly at 95\%~CL, one should use only the $r$-value of the most sensitive analysis. This is however not feasible because for many UL-type results the {\em expected} limits are not publicly available.} 
Points which are not excluded ($r_{\rm max} < 1$) are further studied using the \smodels\ coverage module (see section~3.5 in \cite{Ambrogi:2017neo}).

\begin{table}\centering
\begin{tabular}{llcll}
&{\bf Analysis}              & {\bf Ref.} & {\bf ~~ID} & {\bf SModelS database} \\ \hline
\parbox[t]{3mm}{\multirow{7}{*}{\rotatebox[origin=c]{90}{Inclusive}}} 
&\ZeroLepton                 & \cite{Aad:2014wea} & SUSY-2013-02$^{\,*}$ & 6 UL, 2 EM \\
&\Multijet                   & \cite{Aad:2013wta} & SUSY-2013-04$^{\,*}$ & 1 UL,  10 EM  \\
&\OneLeptonStrong             & \cite{Aad:2015mia} & SUSY-2013-20$^{\,*}$ &  1 UL from CONF-2013-089~\cite{ATLAS-CONF-2013-089}   \\
&\TauStrong                  & \cite{Aad:2014mra} & SUSY-2013-10 &  n.i.        \\
&\SSThreeLepton              & \cite{Aad:2014pda} &  SUSY-2013-09 &  1 UL (+5 UL, CONF-2013-007~\cite{ATLAS-CONF-2013-007})\\  
&\ThreeBjets                 & \cite{Aad:2014lra} &  SUSY-2013-18$^{\,*}$ & 2 UL, 2 EM        \\
&\Monojet                    & \cite{Aad:2015zva} &  ---  &  ---    (but monojet stop, see below)      \\ \hline
\parbox[t]{3mm}{\multirow{7}{*}{\rotatebox[origin=c]{90}{Third generation}}} 
&\ZeroLeptonStop             & \cite{Aad:2014bva} & SUSY-2013-16$^{\,*}$ & 1 UL, 1 EM \\
&\OneLeptonStop              & \cite{Aad:2014kra} & SUSY-2013-15$^{\,*}$ & 1 UL, 1 EM  \\
&\TwoLeptonStop              & \cite{Aad:2014qaa} & SUSY-2013-19$^{\,*}$ & 2 UL         \\
&\StopMonojet                & \cite{Aad:2014nra} &  SUSY-2013-21 &   4 EM       \\
&\StopZ                      & \cite{Aad:2014mha} &  SUSY-2013-08 &  1 UL       \\
&\TwoBjet                    & \cite{Aad:2013ija} &  SUSY-2013-05$^{\,*}$ &  3 UL, 1 EM     \\
&\TBmet                      & \cite{Aad:2015pfx}  &  SUSY-2014-07 &   ---       \\ \hline
\parbox[t]{3mm}{\multirow{6}{*}{\rotatebox[origin=c]{90}{Electroweak}}} 
&\OneLeptonHiggs             & \cite{Aad:2015jqa} & SUSY-2013-23$^{\,*}$ & 1 UL \\
&\TwoLepton                  & \cite{Aad:2014vma} &  SUSY-2013-11 & 4 UL, 4 EM   \\
&\TwoTau                     & \cite{Aad:2014yka} &  SUSY-2013-14 &    ---     \\
&\ThreeLepton                & \cite{Aad:2014nua} & SUSY-2013-12 &  5 UL \\
&\FourLepton                 & \cite{Aad:2014iza} & SUSY-2013-13 &    ---       \\
&\DisappearingTrack          & \cite{Aad:2013yna} &  SUSY-2013-01 &   n.a.\\   \hline
\parbox[t]{3mm}{\multirow{2}{*}{\rotatebox[origin=c]{90}{Other}}} 
&\LLSparticles               & \cite{Aad:2012pra,ATLAS:2014fka} &  --- & n.a.  \\
&\HiggsToTauTau              & \cite{Aad:2014vgg} &   --- & n.a.         \\ \hline
\end{tabular}
\caption{The 22 searches considered in the ATLAS pMSSM study~\cite{Aad:2015baa} and their correspondences in the \smodelsv\ database.  
A superscript $^*$ with the ID means that in addition {\sc Fastlim} EMs for a preliminary version of the analysis are included, see Table~\ref{tab:FastlimAnalyses}. 
The monojet results from \cite{Aad:2015zva} are not implemented in \smodels\ because our SMS assumptions do not apply to dark matter simplified models. 
The analyses \cite{Aad:2015pfx,Aad:2014yka,Aad:2014iza} do not provide useable SMS interpretations. Finally, searches for new resonances, long-lived particles, and disappearing tracks \cite{Aad:2013yna,Aad:2012pra,ATLAS:2014fka,Aad:2014vgg} currently cannot be treated in the \smodels\ framework. }
\label{tab:AtlasSearchesInPaper}
\end{table}

\begin{table}\centering
\begin{tabular}{llcc}
&{\bf Analysis}              & {\bf Ref.} & {\bf ID}  \\ \hline
\parbox[t]{3mm}{\multirow{4}{*}{\rotatebox[origin=c]{90}{Incl.}}} 
& \ZeroLepton & \cite{ATLAS-CONF-2013-047} & ATLAS-CONF-2013-047 \\ 
& \Multijet & \cite{ATLAS-CONF-2013-054} & ATLAS-CONF-2013-054 \\ 
& \OneLeptonStrong & \cite{ATLAS-CONF-2013-062} & ATLAS-CONF-2013-062 \\ 
& \ThreeBjets & \cite{ATLAS-CONF-2013-061} & ATLAS-CONF-2013-061 \\ 
\hline
\parbox[t]{3mm}{\multirow{4}{*}{\rotatebox[origin=c]{90}{Third gen.}}} 
& \ZeroLeptonStop & \cite{ATLAS-CONF-2013-024} & ATLAS-CONF-2013-024 \\ 
& \OneLeptonStop & \cite{ATLAS-CONF-2013-037} & ATLAS-CONF-2013-037 \\ 
& \TwoLeptonStop & \cite{ATLAS-CONF-2013-048} & ATLAS-CONF-2013-048 \\ 
& \TwoBjet & \cite{ATLAS-CONF-2013-053} & ATLAS-CONF-2013-053 \\ 
\hline
\parbox[t]{3mm}{\rotatebox[origin=c]{90}{EW}}
& \OneLeptonHiggs & \cite{ATLAS-CONF-2013-093} & ATLAS-CONF-2013-093 \\ 
\hline
\end{tabular}
\caption{Analyses covered by the \fastlim~\cite{Papucci:2014rja} EMs converted to the \smodels\ format. 
For each analysis, \fastlim\ considers 24 topologies covering stop-, sbottom- and gluino-pair production 
with direct or cascade decays to a higgsino LSP, inspired by the idea of ``natural SUSY''. 
As explained in the text, efficiencies with uncertainties $>25\%$ are set to zero, so in practice we use 163 of the \fastlim\ EMs.}
\label{tab:FastlimAnalyses}
\end{table}

\begin{table}\centering
\begin{tabular}{llcll}
&{\bf Analysis}              & {\bf Ref.} & {\bf ~~ID} & {\bf SModelS database} \\ \hline
\parbox[t]{3mm}{\multirow{9}{*}{\rotatebox[origin=c]{90}{Gluino, Squark}}} 
& jets + \met, $\alpha_T$   & \cite{Chatrchyan:1527115} & SUS-12-028 & 4 UL \\
& 3(1$b$-)jets + \met  & \cite{Chatrchyan:1546693} & SUS-12-024 & 2 UL, 3 EM \\
& jet multiplicity + $H_T^{\rm miss}$  & \cite{Chatrchyan:1662652} & SUS-13-012 & 4 UL, 20 EM \\
& $\ge 2$ jets + \met, $M_{T2}$  & \cite{Khachatryan:1989788} & SUS-13-019 & 8 UL \\
& $\ge 1b$ + \met, Razor  & \cite{Khachatryan:1984165} & SUS-13-004 & 5 UL \\
& 1 lepton + $\ge 2b$-jets + \met  & \cite{CMS-PAS-SUS-13-007} & SUS-13-007 & 3 UL, 2 EM \\
& 2 OS lept. + $\ge$4(2$b$-)jets + \met  & \cite{CMS-PAS-SUS-13-016} & PAS-SUS-13-016 & 2 UL \\
& 2 SS leptons + $b$-jets + \met  & \cite{Chatrchyan:1631468} & SUS-13-013 & 4 UL, 2 EM \\
& $b$-jets + 4 $W$s + \met  & \cite{Khachatryan:1976453} & SUS-14-010 & 2 UL \\
\hline
\parbox[t]{3mm}{\multirow{5}{*}{\rotatebox[origin=c]{90}{Third gen.}}} 
&0 lepton + $\ge5$(1$b$-)jets + \met  & \cite{CMS-PAS-SUS-13-015} & PAS-SUS-13-015 & 2 EM \\
&0 lepton + $\ge6$(1$b$-)jets + \met  & \cite{CMS-PAS-SUS-13-023} & PAS-SUS-13-023 & 4 UL \\
&1 lepton + $\ge4$(1$b$-)jets + \met  & \cite{Chatrchyan:1567175} & SUS-13-011 & 4 UL, 2 EM \\
& $b$-jets + \met  & \cite{CMS-PAS-SUS-13-018} & PAS-SUS-13-018 & 1 UL \\
& soft leptons, few jets + \met  & \cite{Khachatryan:2117955} & SUS-14-021 & 2 UL \\
\hline
\parbox[t]{3mm}{\rotatebox[origin=c]{90}{EW}}
&multi-leptons + \met             & \cite{Khachatryan:1704963} & SUS-13-006 & 6 UL \\
\hline
\end{tabular}
\caption{CMS 8~TeV results included in the \smodelsv\ database and used in addition to the ATLAS 
results in Tables~\ref{tab:AtlasSearchesInPaper} and \ref{tab:FastlimAnalyses}.}
\label{tab:CMSAnalyses}
\end{table}

\clearpage

\section{Exclusion compared to ATLAS}\label{sec:results}

As a first overview of our results, we list in Table~\ref{tab:smodpmssmsum} the total number of points studied, 
the number of points that can be excluded by \smodels\ ($r_{\rm max}\ge1$) when using only the UL results in the database, 
and the number of points that can be excluded when using the full 8~TeV database, that is including EM results.
We see that the coverage of bino-like LSP scenarios can be improved by using EMs,
increasing from 44\% (UL results only) to 55\% (full database).
Similarly, the coverage for the higgsino-like LSP scenarios is improved from 55\% to 63\%. 

\begin{table}[h!]
\centering
\begin{tabular}{c | c | c}
 & Bino-like LSP & Higgsino-like LSP\\
\hline
Total number of points & 38575 & 45594 \\
\hline
Number of points excluded -- UL results only & 16957 & 25024 \\
\hline
Number of points excuded -- full database & 21151 & 28669
\end{tabular}
\caption{Summary of results, listing the number of ATLAS-excluded pMSSM points tested in this study, the number of points excluded by \smodels\ when using UL-type results only, 
and the number of points excluded when using the full 8 TeV database including EM-type results.}
\label{tab:smodpmssmsum}
\end{table}

\begin{figure}[ht!]\centering
\includegraphics[width=0.5\textwidth]{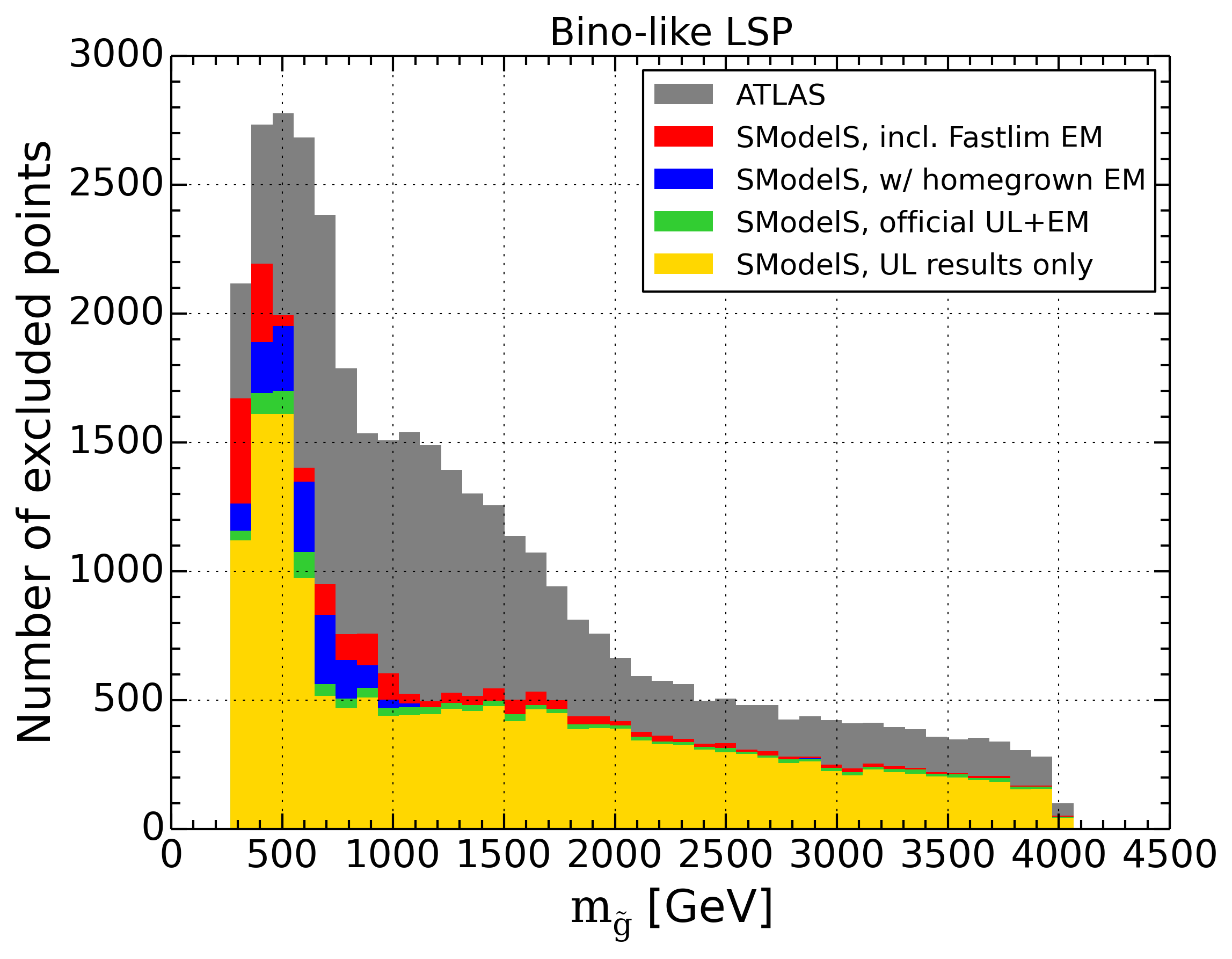}%
\includegraphics[width=0.5\textwidth]{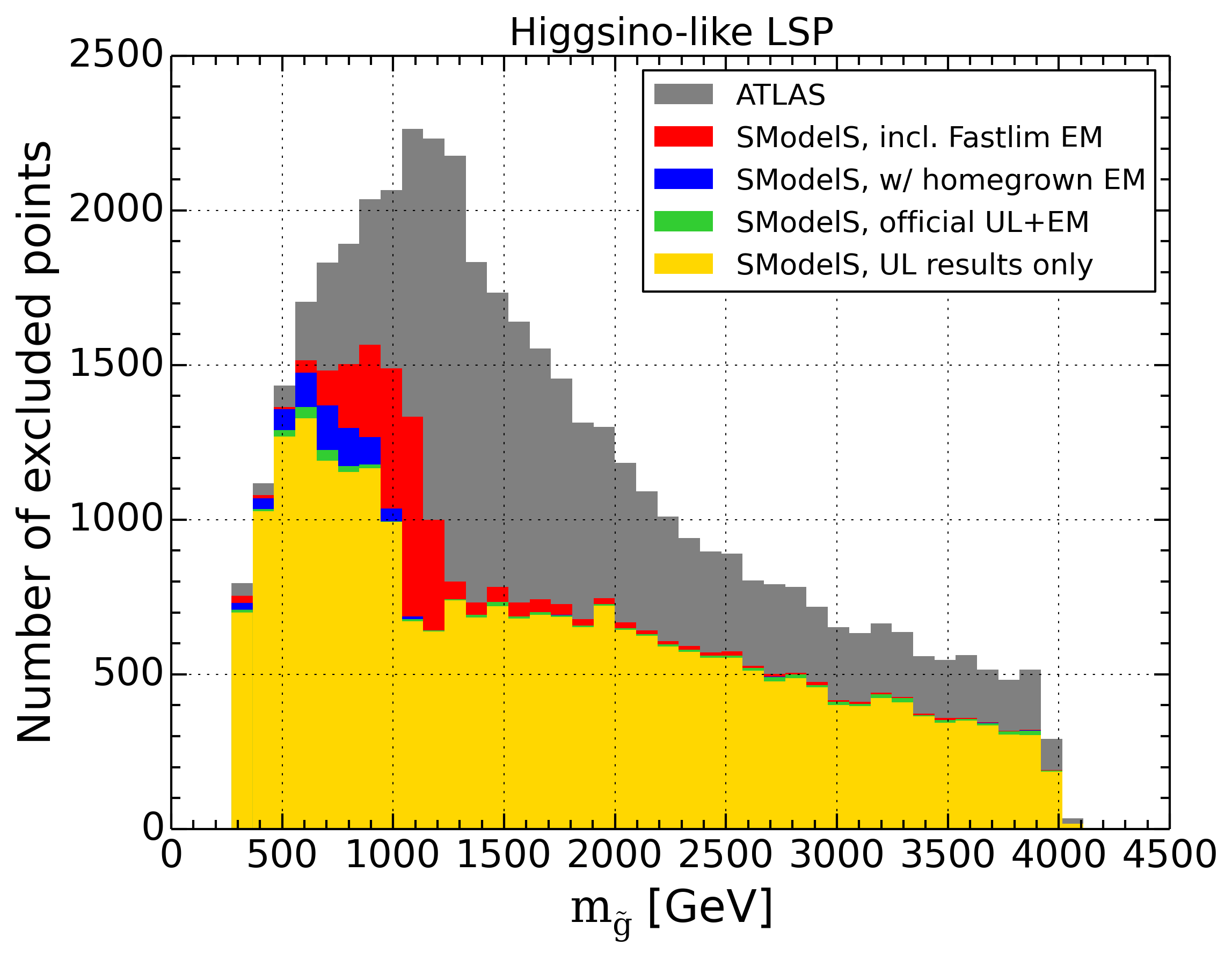}
\caption{Number of points excluded by \smodels\  using only UL results (in yellow), adding official EM results (in green), adding `home-grown'  EMs (in blue) and finally adding also \fastlim\ EMs (in red). For reference the total number of ATLAS-excluded points is also shown (in grey). On the left for bino-like LSP and on the right for higgsino-like LSP.
\label{fig:pmssmGlu1d}}
\end{figure}

\noindent
The improvement in coverage due to EMs largely happens for light to intermediate gluino masses,
as illustrated in Figure~\ref{fig:pmssmGlu1d}. These scenarios benefit
from the fact that EMs allow us to combine the signal for all topologies contributing to the same signal region before comparing
against an overall cross section limit, hence increasing the constraining power.
Moreover, some asymmetric topologies are included in the EM-type results (from \fastlim) 
but not in the UL-type results in the database.
Figure~\ref{fig:pmssmGlu1d} also shows the importance of the \fastlim\ and our `home-grown' EMs with respect to the official ATLAS and CMS SMS results. We note that the \fastlim\ maps are particularly relevant for constraining 
gluinos in the intermediate mass range decaying to higgsino-like EW-inos, which is typical for the natural SUSY case they have been derived for. 
In numbers, official UL and EM results exclude 46\% (56\%) of the bino-LSP (higgsino-LSP) points,
which improves to 50\% (57\%) when adding our `home-grown' EMs, and to the above-mentioned 55\% (63\%) when including in addition \fastlim\ results. 
In the following, we discuss in some detail why still a large fraction of points escapes exclusion by SMS results and how the coverage could be improved. 

\subsection{Gluinos}\label{sec:gluinos}

\begin{figure}[t!]\centering
\includegraphics[width=0.5\textwidth]{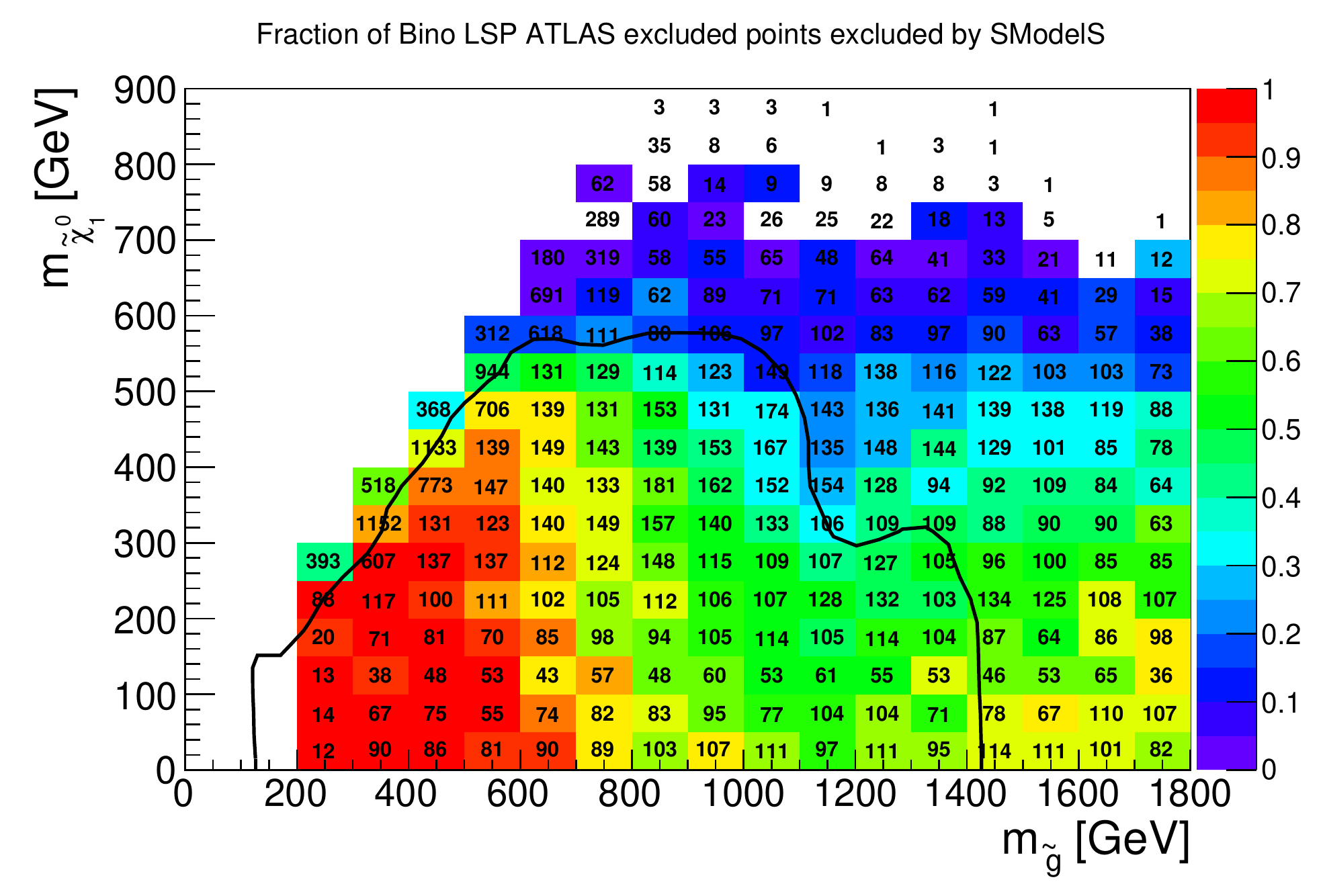}%
\includegraphics[width=0.5\textwidth]{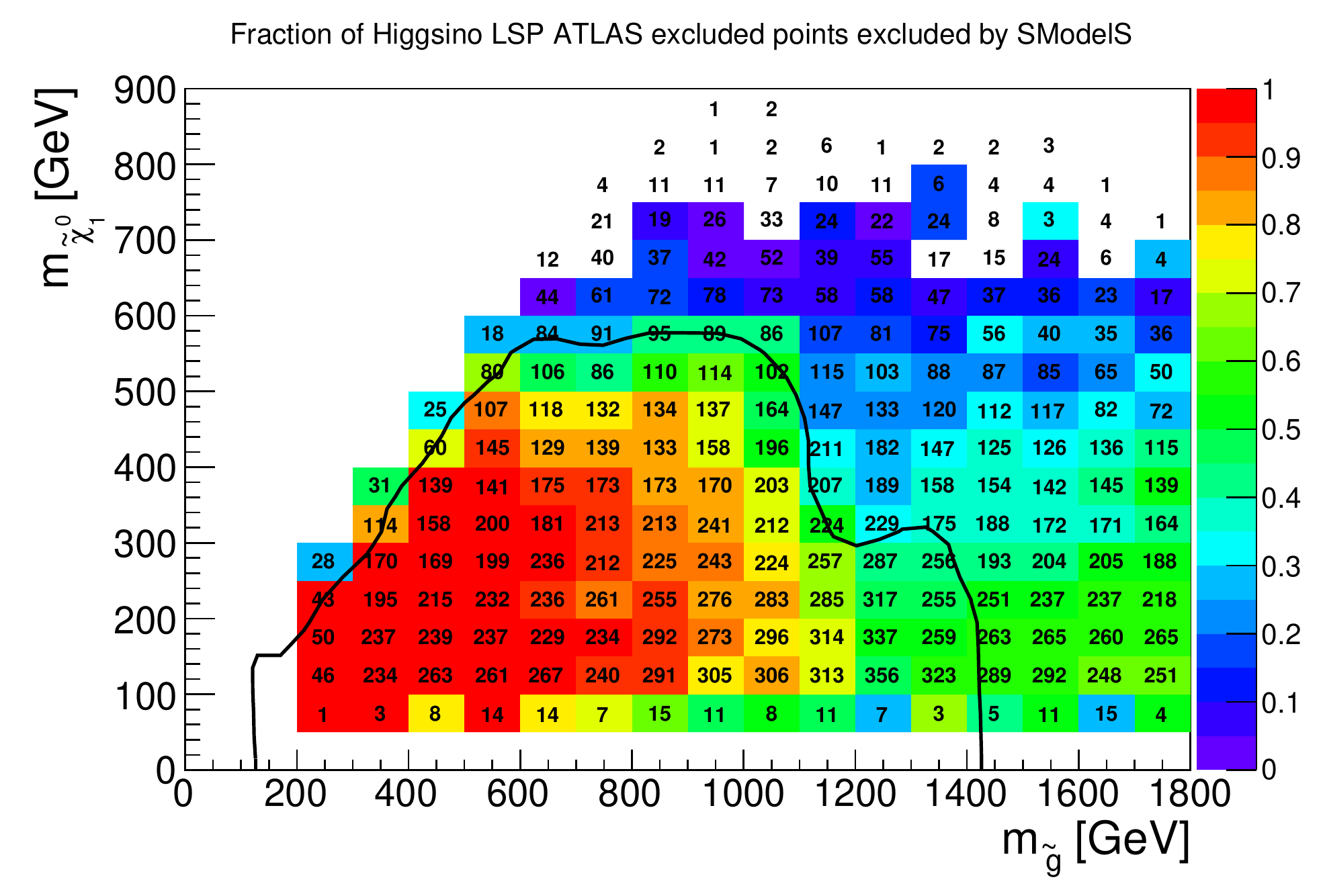}
\caption{Coverage in the gluino vs.\ neutralino mass plane, for gluino masses up to $2$~TeV, for bino-like LSP scenarios (left) and higgsino-like LSP scenarios (right).
The color code indicates the fraction of points excluded by \smodels, the
text gives the total number of points tested in each bin. 
For comparison, the 95\%~CL exclusion line for the $\tilde{g}\rightarrow q q \tilde{\chi}^0_1$ simplified model from~\cite{Aad:2014wea} is drawn in black.
\label{fig:pmssmGlu2d}}
\end{figure}

It is striking that there are many points with light gluinos which cannot be excluded by the SMS results in the \smodels\ database.
To understand this better we show in Figure~\ref{fig:pmssmGlu2d} the coverage in the gluino vs.\ neutralino mass plane. 
For comparison with the ``naive'' SMS expectation, the exclusion line obtained in~\cite{Aad:2014wea} for a simplified model where 
pair-produced gluinos decay exclusively as
$\tilde{g}\rightarrow q q \tilde{\chi}^0_1$ is also drawn in Figure~\ref{fig:pmssmGlu2d}. 
We see that light gluinos escape SMS limits especially in the compressed region where monojet-type searches become important. This is in agreement with the 
simplified model exclusion line. 
Moreover, while the coverage is good for very light gluinos up to about 600~GeV,   
it drops for intermediate gluino masses around $1$~TeV and higher, 
as can also be observed in Figure~\ref{fig:pmssmGlu1d}.
This is particularly pronounced in the bino-like LSP scenario.
Concretely, the coverage of bino-like LSP scenarios is 80\% when considering only points with light gluinos ($m_{\tilde{g}} < 600$~GeV), but drops
to 60\% when considering all points with $m_{\tilde{g}} < 1400$~GeV.
Similarly the coverage of higgsino-like LSP scenarios drops from 97\% ($m_{\tilde{g}} < 600$~GeV) to 74\% ($m_{\tilde{g}} < 1400$~GeV).
Note that for bino-like LSP scenarios light gluinos are mainly found in the compressed region ($m_{\tilde{g}} - m_{\tilde{\chi}^0_1} < 100$~GeV),
where the bins contain a large number of model points. This is not the case for higgsino-like LSP scenarios.

The somewhat better coverage of non-compressed sub-TeV gluinos in the higgsino-like LSP set can be understood as follows.
In the case of a bino-like LSP, unless the gluino-LSP mass difference is small,
direct decays into the LSP often have only 30\% or less branching ratio.
Decays into wino- or higgsino-like states are often more important, 
leading to cascade decays into the LSP and to asymmetric branches with different final states and, possibly, different intermediate masses.%
\footnote{Asymmetric branches can occur from pair production when the two initially produced SUSY particles undergo different decays, or from associated production of two different SUSY particles.} 
This reduces the fraction of gluino signatures covered by SMS results, and as the total cross section reduces with increasing gluino mass,
the fraction that can be constrained is no longer large enough to exclude the point.
For higgsino-like LSP scenarios, on the other hand, 
the second neutralino $\tilde{\chi}_2^0$ as well as the lighter chargino $\tilde{\chi}_1^{\pm}$ are
nearly degenerate with the LSP, and their decay can often be mass compressed in \smodels.
In this case, contributions from $\tilde g\to qq'\tilde\chi^\pm_1$, $qq\tilde\chi^0_2$ and $qq\tilde\chi^0_1$ can be summed up, 
which explains the better coverage of light gluinos in the higgsino-LSP case already by UL results seen in Figure~\ref{fig:pmssmGlu1d}. 
Moreover, gluino decays into third generation are often dominant in the higgsino-LSP case, leading to a mix of final states ($4b$, $4t$, $2b2t$, $3b1t$, $1b3t$) which can in part be covered by the \fastlim\ EMs. 

\begin{figure}[t!]\centering
\includegraphics[width=0.5\textwidth]{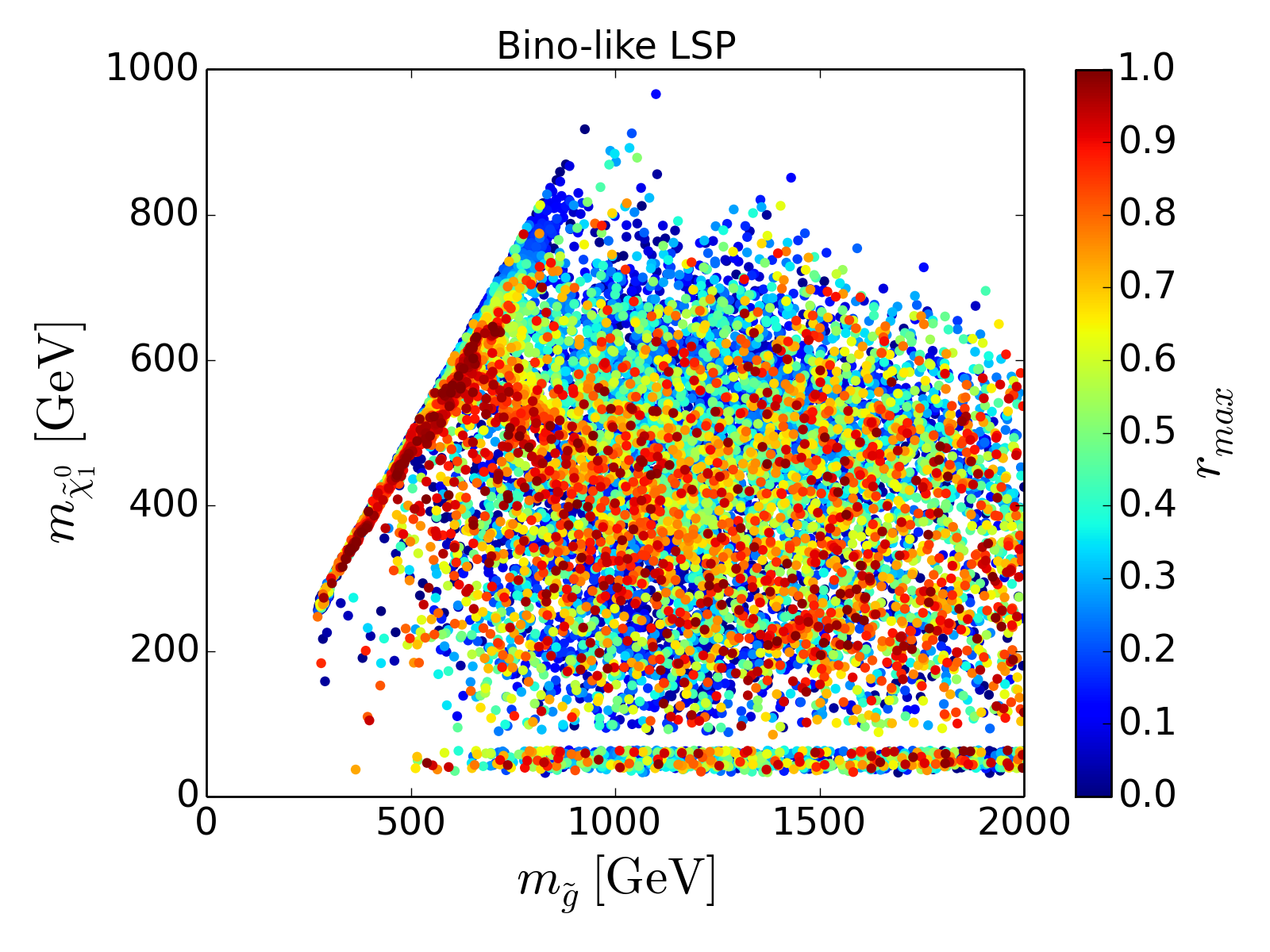}%
\includegraphics[width=0.5\textwidth]{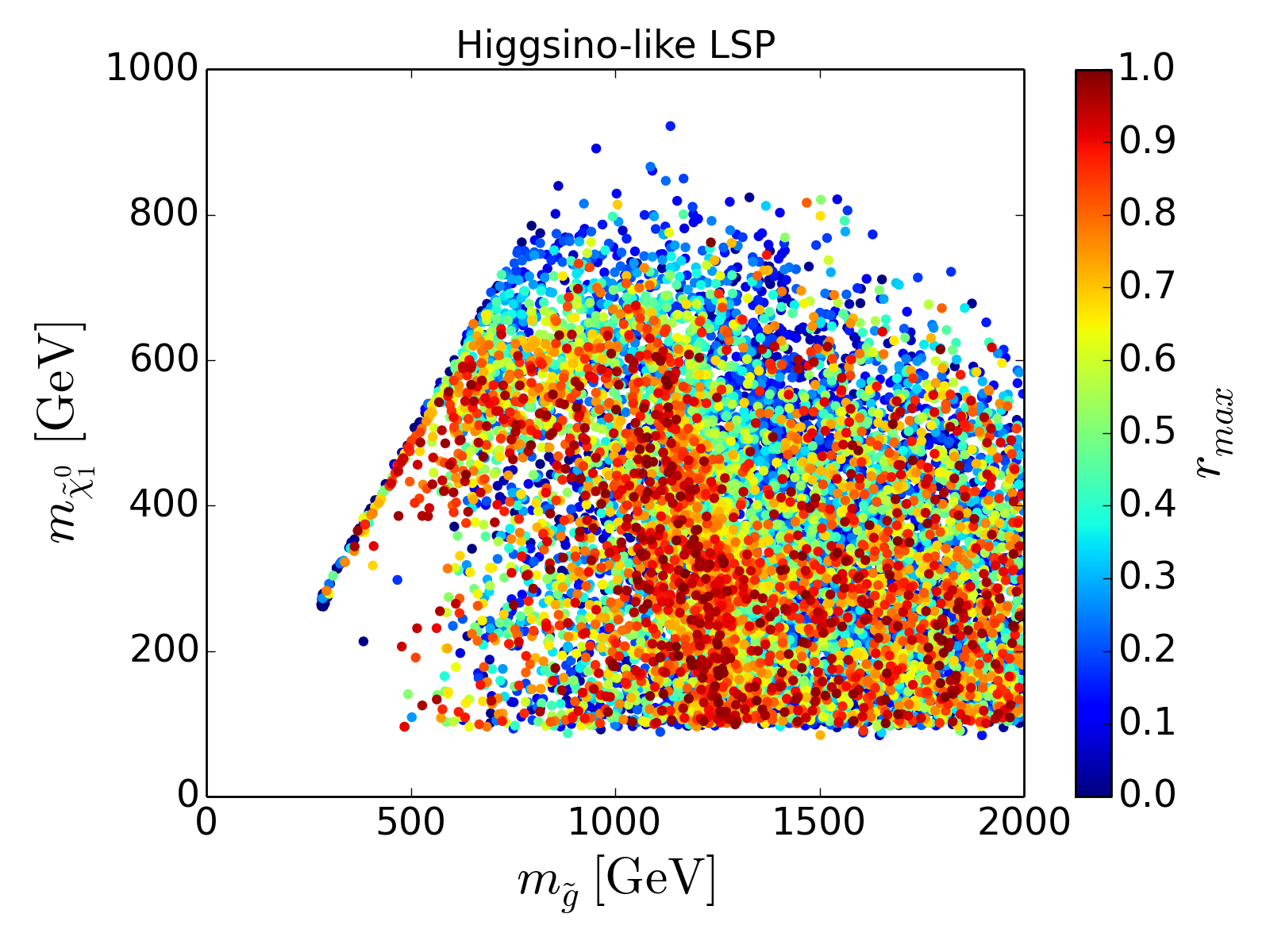}
\caption{Maximum $r$ value reported by \smodels\ for allowed points, for gluino masses up to $2$~TeV, for bino-like LSP scenarios (left) and higgsino-like LSP scenarios (right). Points are sorted from low to high $r$-values, with the highest values of $r$ shown on top.
\label{fig:pmssmGluRval}}
\end{figure}

Another important consideration is how far the points which escape the \smodels\ exclusion are from becoming excluded.
Uncertainties inherent to the \smodels\ approach and the fact that we used LO cross-sections
for EW process (while ATLAS used NLO values) can reduce the
exclusion reach.
In Figure~\ref{fig:pmssmGluRval} we show the maximum $r$ values found for points escaping exclusion by \smodels.
We see that many points, especially in the region of intermediate gluino masses and in the more compressed region, are in fact
close to the exclusion limit.
We therefore expect that the coverage can be considerably improved by adding additional EMs, thus allowing to
test a larger fraction of the total cross section.
Furthermore, we find that 10\% of bino-like LSP scenarios and 12\% of higgsino-like LSP scenarios have $0.8<r_{\rm max}<1.2$,
which allows a rough estimate of the uncertainties involved in the exclusion.
(The overall systematic uncertainty is estimated to be of the level of 20\%~\cite{Ambrogi:2017neo}.)
In turn, we find $r_{\rm max}>1.2$ for 50\% of bino-like LSP and 58\% of higgsino-like LSP scenarios.

\begin{figure}[t!]\centering
\includegraphics[width=0.5\textwidth]{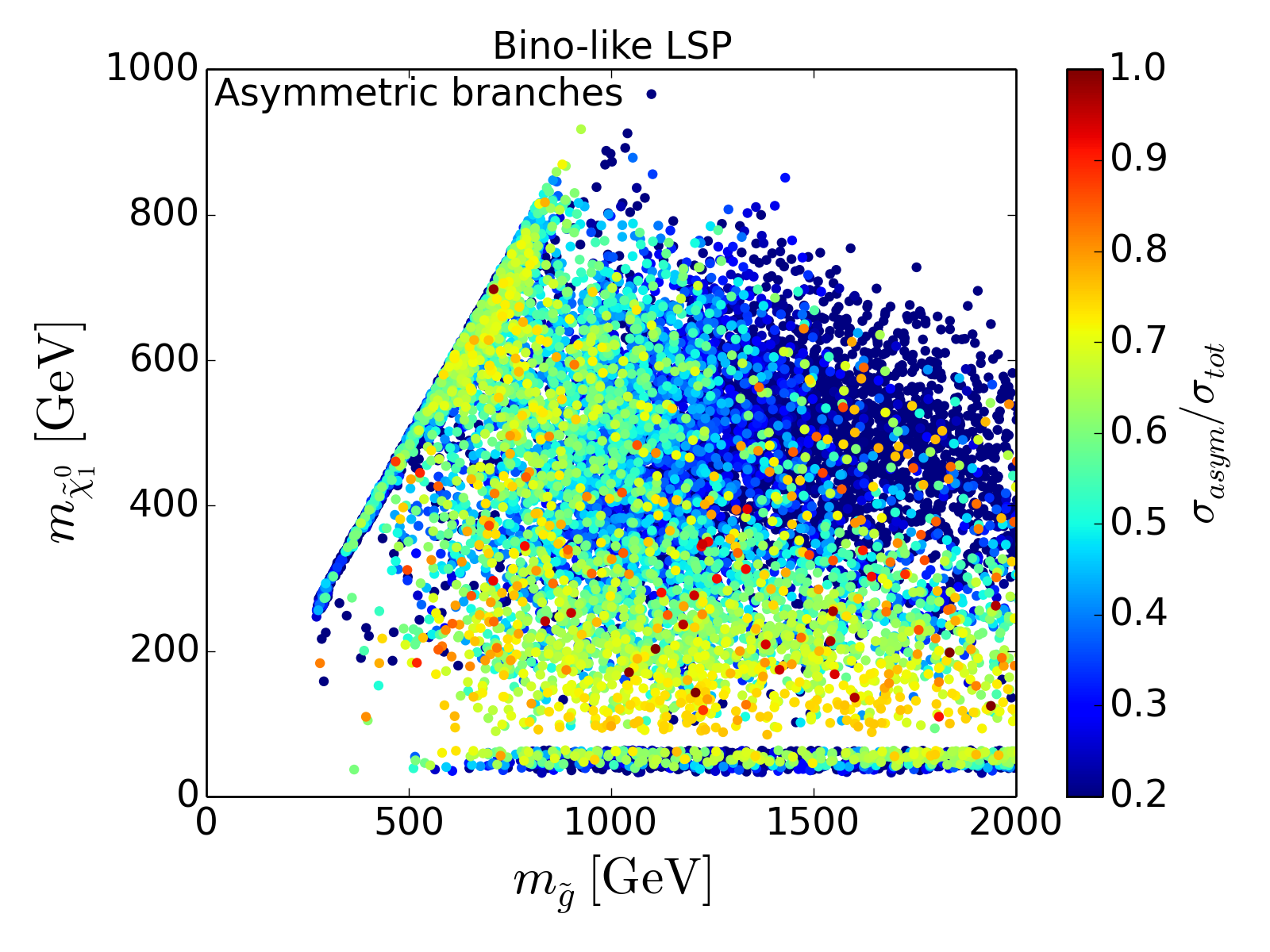}%
\includegraphics[width=0.5\textwidth]{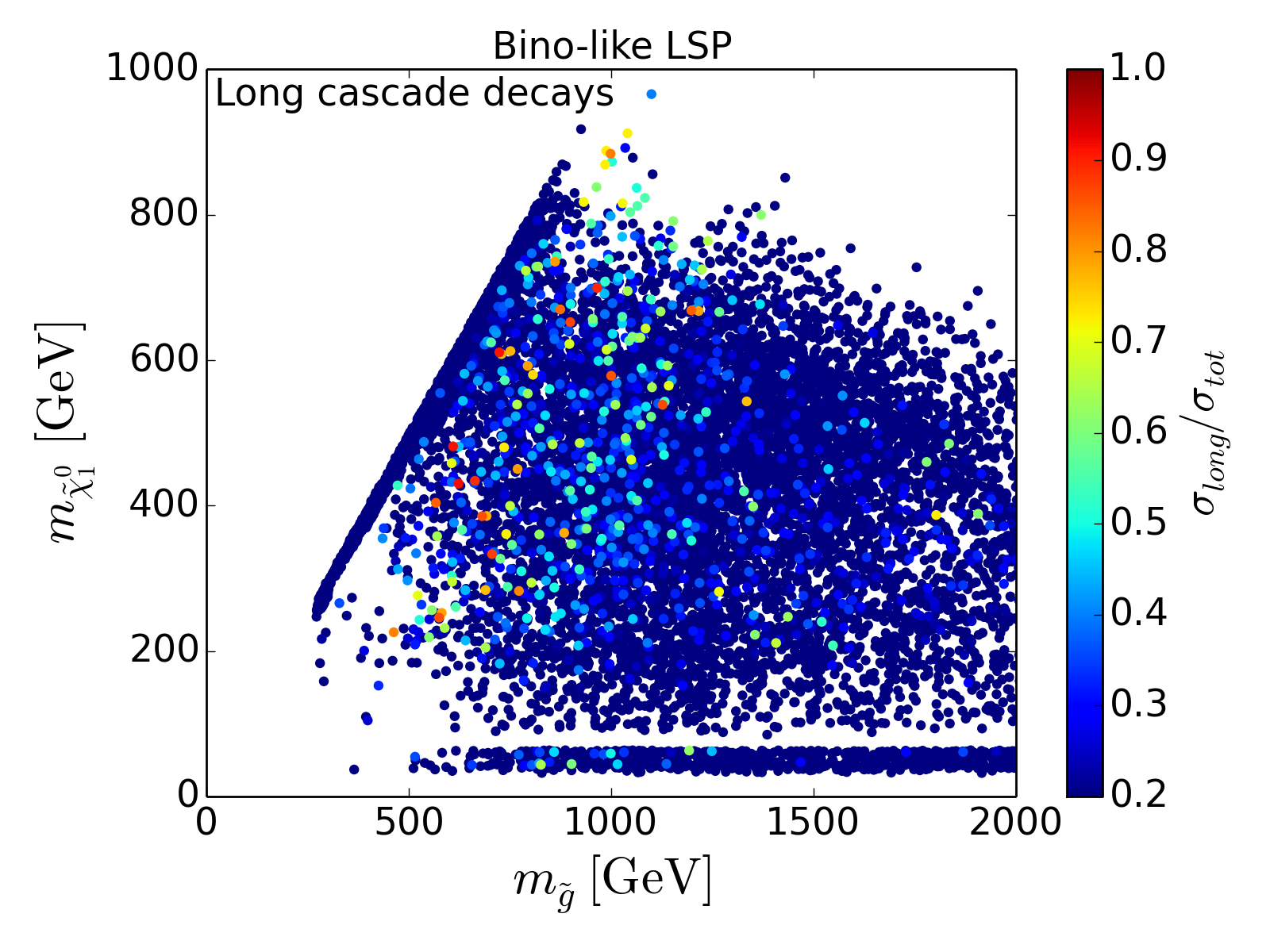}
\caption{Relative cross section in unconstrained decays with asymmetric branches (left) and long cascade decays (right), for scenarios with a
bino-like LSP. Here the total cross section $\sigma_{tot}$ refers to the full $8$~TeV SUSY cross section.
Only \smodels-allowed points with total cross section larger than $10$~fb are considered.
\label{fig:pmssmBinoAsym}}
\end{figure}

\begin{figure}[t!]\centering
\includegraphics[width=0.5\textwidth]{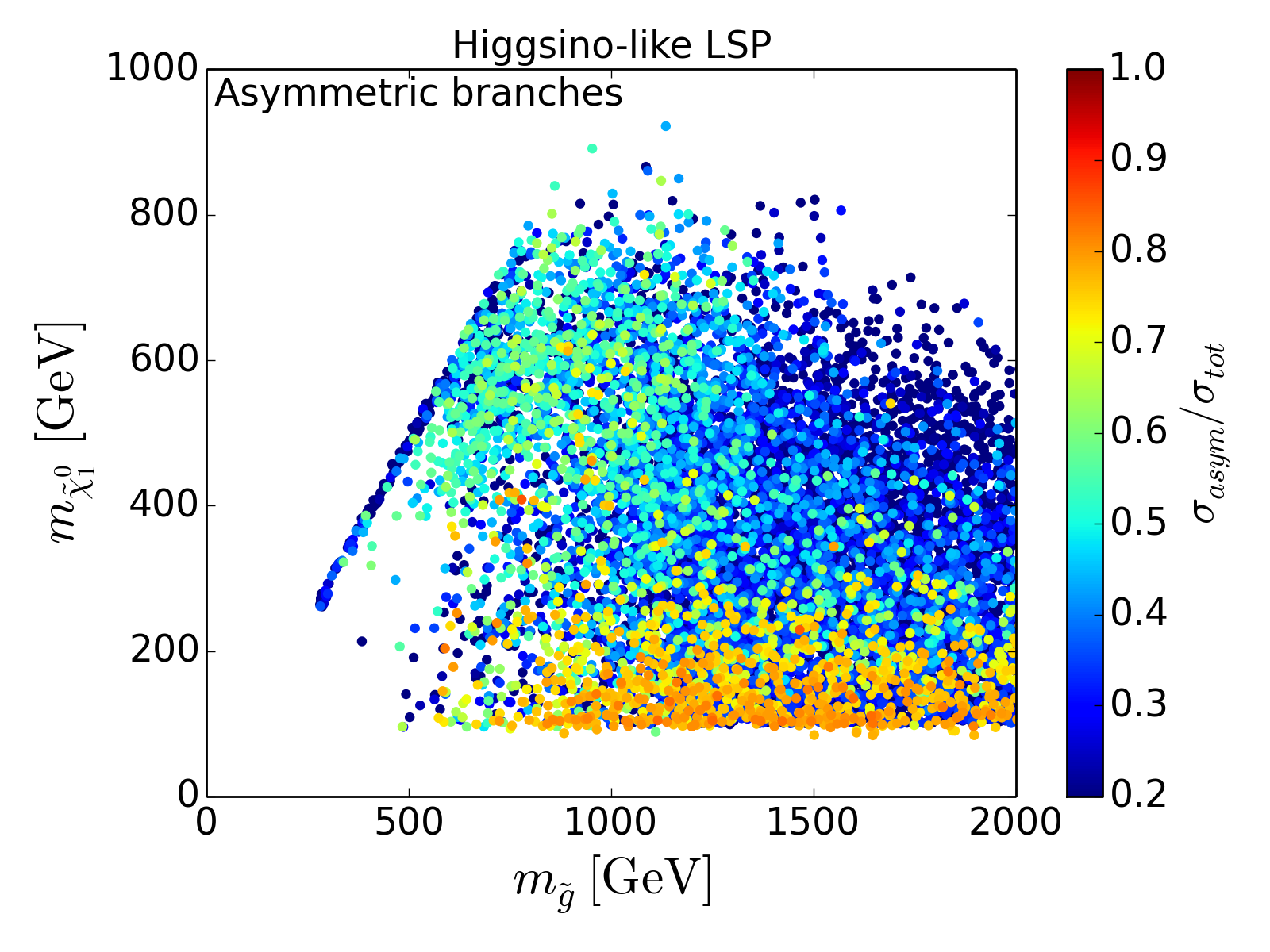}%
\includegraphics[width=0.5\textwidth]{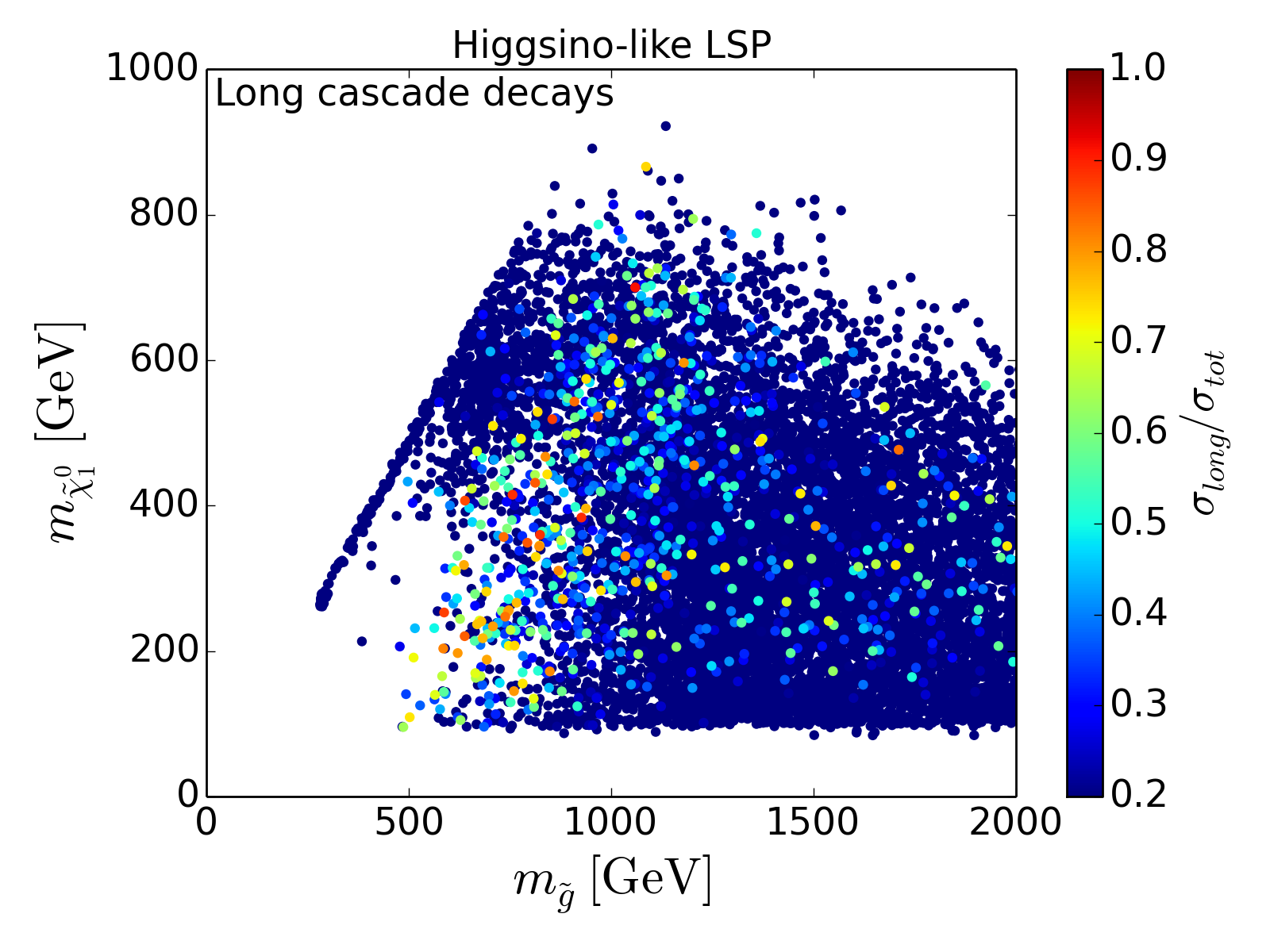}
\caption{Same as Figure~\ref{fig:pmssmBinoAsym} but for points with a higgsino-like LSP.
\label{fig:pmssmHiggsinoAsym}}
\end{figure}

To understand the possibilities of further improving the coverage, without going into details
about the specific missing topologies,\footnote{In \smodelsnn, ``missing topologies'' are defined as topologies for which no experimental result is available in the database.} we show in Figures~\ref{fig:pmssmBinoAsym} and \ref{fig:pmssmHiggsinoAsym}
the relative cross sections of \smodels-allowed\footnote{We define ``\smodels-allowed'' as ``excluded by ATLAS but not excluded by \smodels''.} 
points which go into missing topologies with asymmetric branches (left) or long cascade decays (right), for
bino-like LSP scenarios and higgsino-like LSP scenarios respectively. 
In this classification, asymmetric branch topologies have at most one intermediate odd particle in each
branch, so that the number of new particles and mass parameters still is sufficiently small for a viable SMS interpretation. 
On the other hand, as long cascade decays we define decay chains with two or more intermediate odd particles 
and we no longer consider a simplified model description viable. 
We see that in fact topologies with asymmetric decay branches are important for a large number of points for both bino- and higgsino-like LSP scenarios,
whereas long cascade decay topologies are dominant only in a few cases. 
Therefore inclusion of additional asymmetric topologies should have a significant impact on the SMS coverage.

\begin{figure}[t!]\centering
\includegraphics[width=0.5\textwidth]{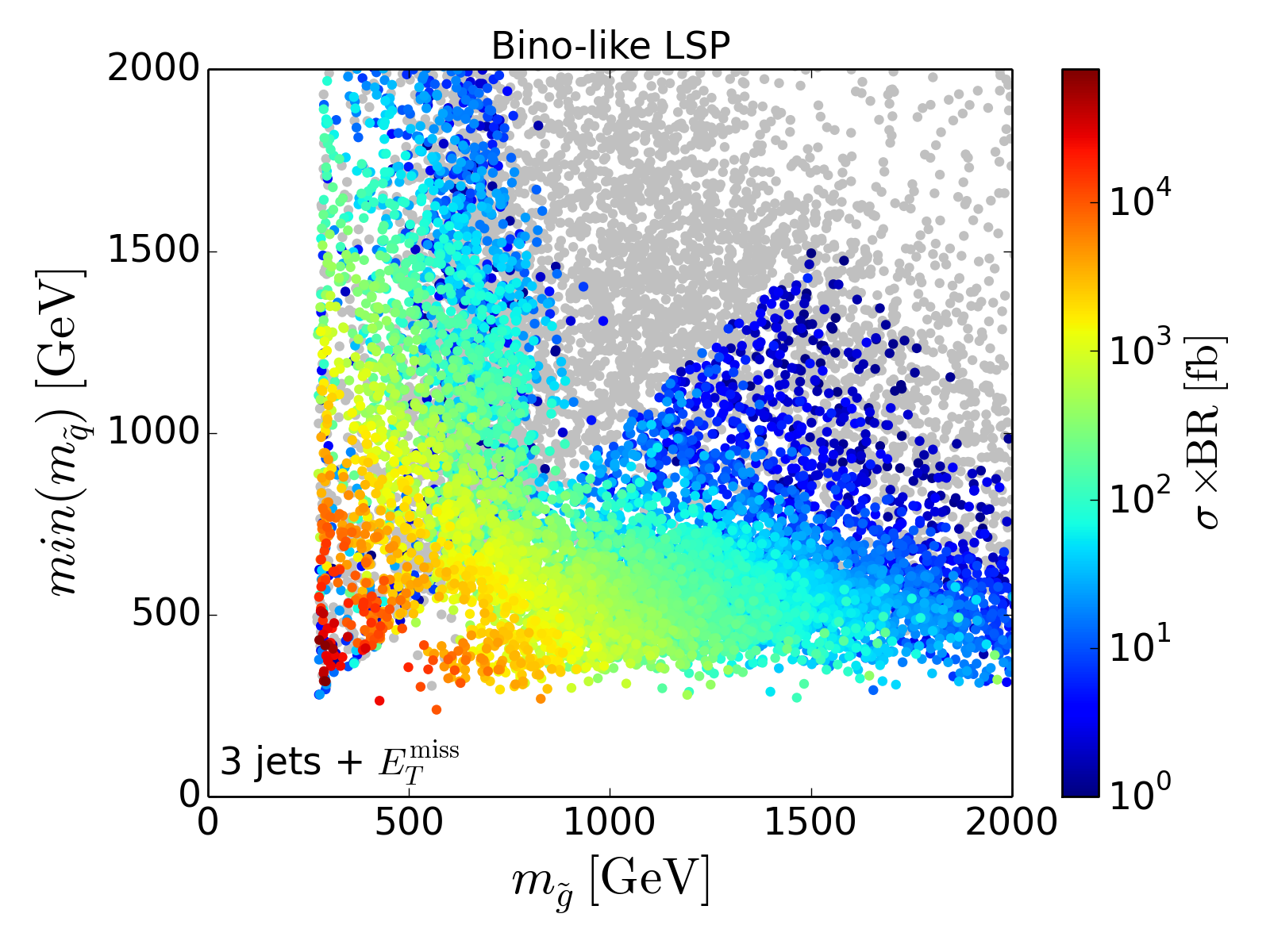}%
\includegraphics[width=0.5\textwidth]{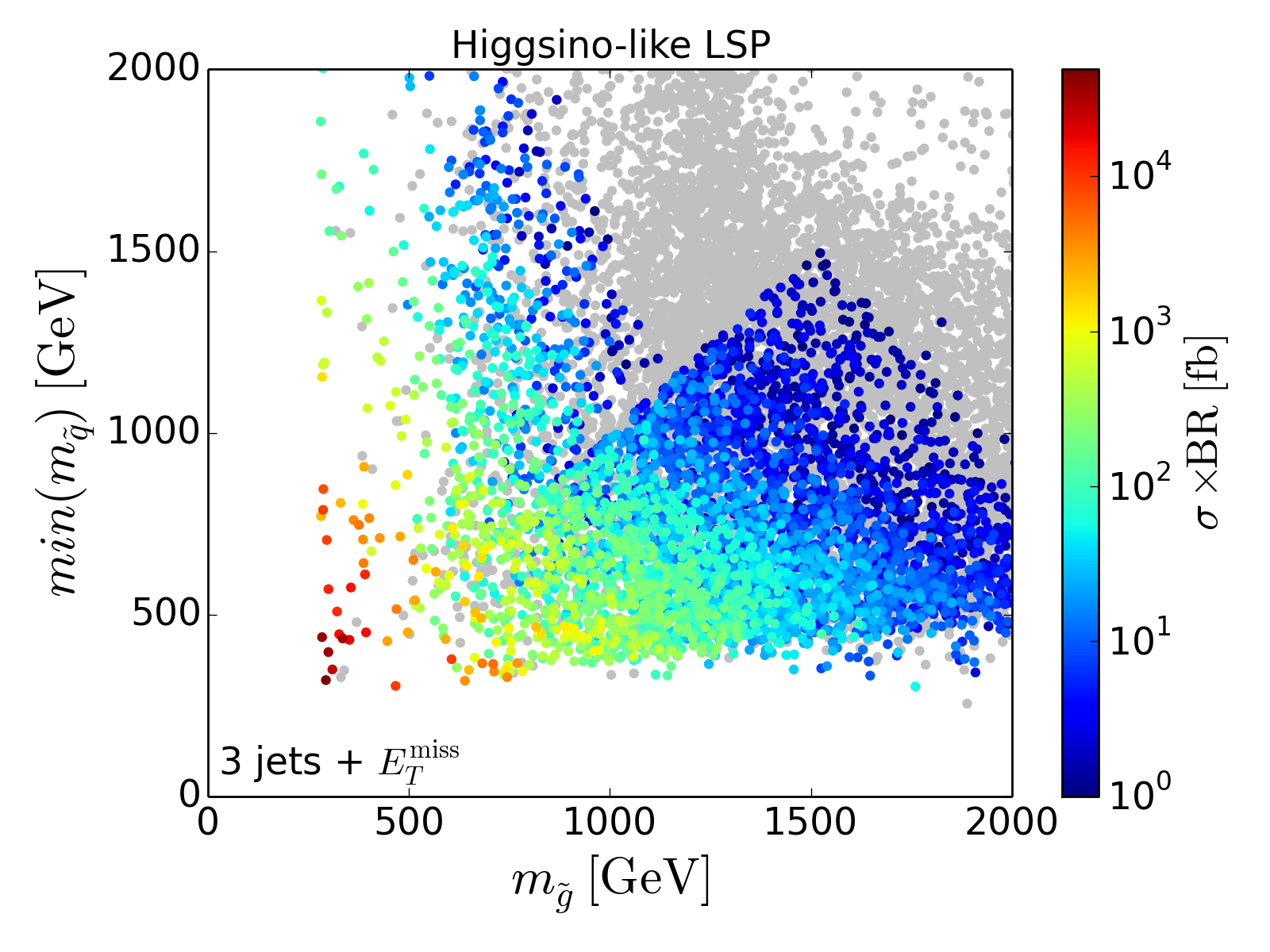}
\caption{Cross section for the $\tilde g\tilde q\to 3$~jets + $\MET$ missing topology in the gluino vs.\ squark mass plane, for bino-like LSP (left) and higgsino-like
LSP (right). Only \smodels-allowed points are considered.
\label{fig:pmssmGluSq}}
\end{figure}

A particularly important missing topology with asymmetric branches arises from gluino-squark associate production, giving
a 3~jets + $\MET$ final state. This is important in particular when the light-flavor squarks are highly split and the gluino can 
decay to a single on-shell squark. 
The relevant process is $pp\to \tilde g\tilde q$ followed by $\tilde q\to q\tilde\chi^0_1$ on one branch and $\tilde g\to q\tilde q\to q\bar q \tilde\chi^0_1$ on the other branch.
The same topology is also possible when gluinos are lighter than all squarks and decay dominantly via a loop decay to a gluon and the neutralino LSP. 
In this case we have $pp\to \tilde g\tilde q$ followed by $\tilde g\to g\tilde\chi^0_1$ on one branch and $\tilde q\to q\tilde g\to qg \tilde\chi^0_1$ on the other. 
Figure~\ref{fig:pmssmGluSq} shows the cross section of this 
topology in the plane of gluinos mass versus mass of the lightest squark. 
Note that searches for gluino-squark production are typically interpreted either in a simplified model where gluinos and squarks are (nearly) mass-degenerate, or in a minimal gluino-squark model where all production processes---gluino pairs, squark pairs, and gluino-squark associated production---are combined~\cite{Aad:2014wea}.
Such results cannot be used for reinterpretation in generic scenarios where typically the gluino mass differs from the squark masses, and
where the relative importance of the various production and decay channels will be different from the minimal gluino-squark model description.

\begin{figure}[t!]\centering
\includegraphics[width=0.5\textwidth]{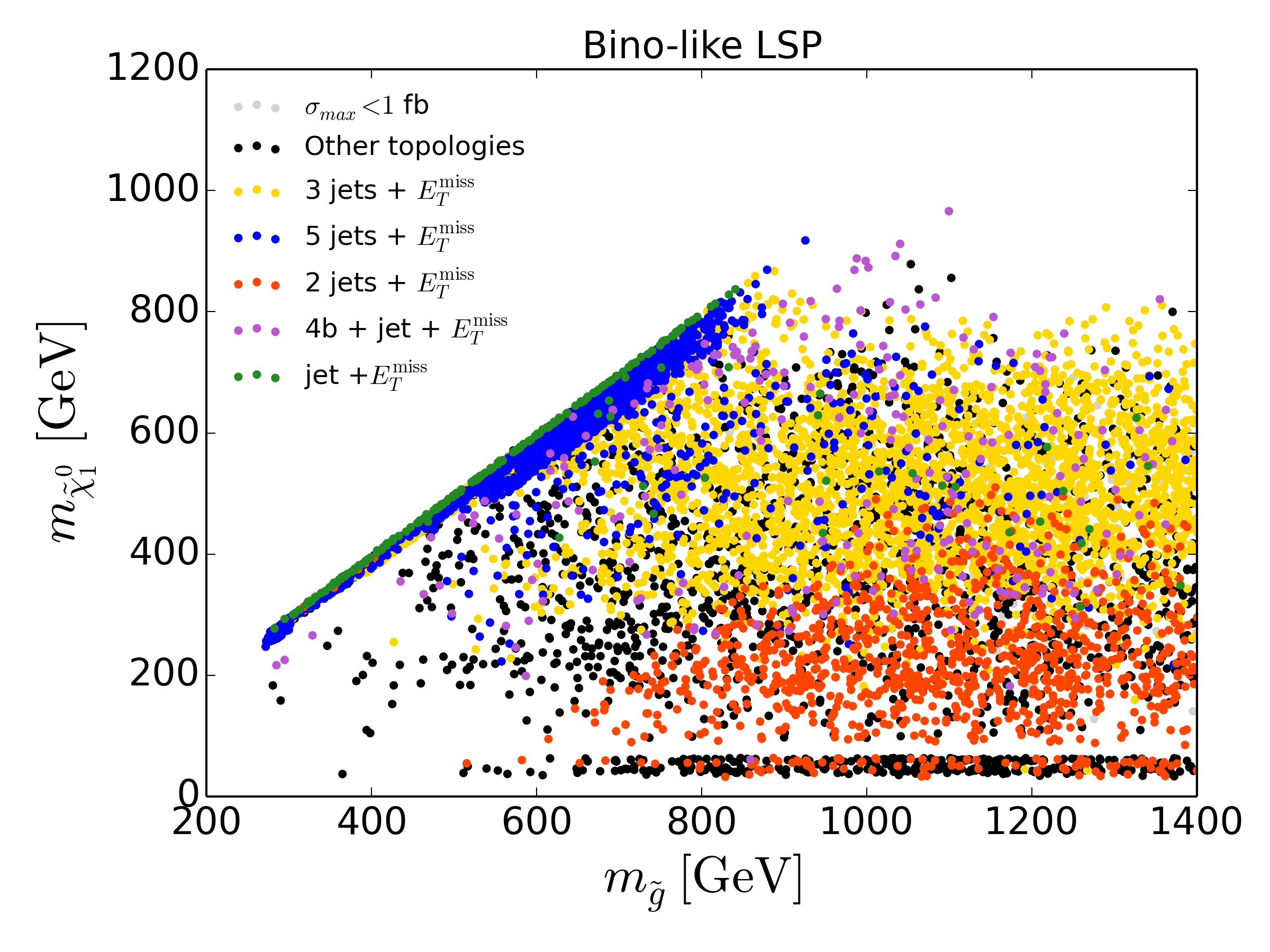}%
\includegraphics[width=0.5\textwidth]{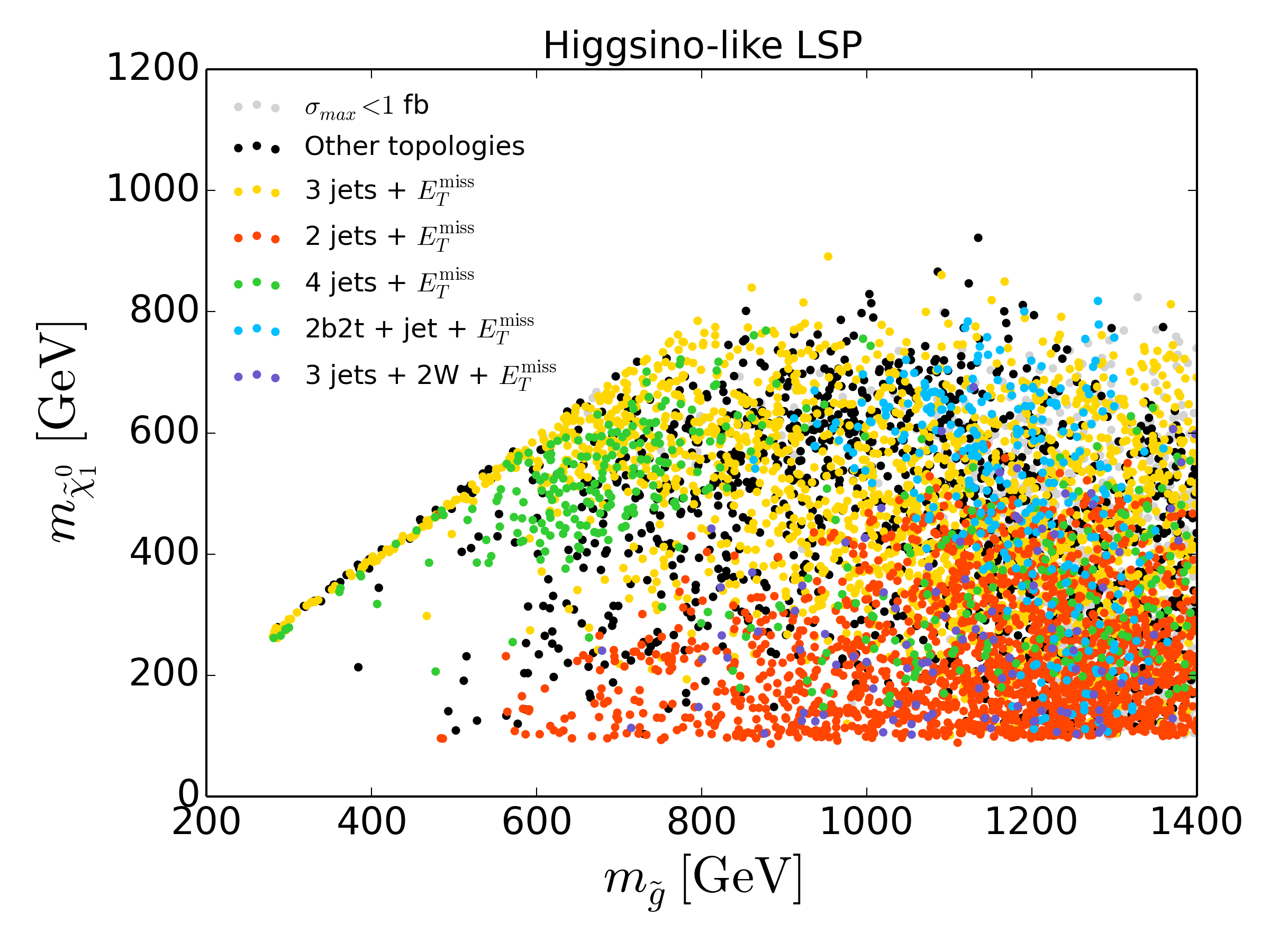}
\caption{Most important missing topologies for \smodels-allowed points with light gluinos. 
The legend lists, from top to bottom, the missing topologies with highest cross sections ordered by their by frequency of occurrence (points in color).
The relevant diagrams, SUSY processes and labelling in \smodels\ notation are given in Appendix~\ref{sec:missingtopotable}. 
\label{fig:GluMissingTopo}}
\end{figure}

The importance of the 
3~jets + $\MET$ topology is corroborated in Figure~\ref{fig:GluMissingTopo}, which shows 
the five most important missing topologies for points with light gluinos below $1.5$~TeV.%
\footnote{For this classification, we first select for each allowed point the missing topology with the highest cross section. These are then 
sorted by frequency of occurrence in the mass range considered.}  
The leading missing topology for both the bino- and the higgsino-LSP datasets is indeed 3~jets + $\MET$ from gluino-squark associated production as discussed above, see the yellow points in Figure~\ref{fig:GluMissingTopo} which cover a wide range of gluino and LSP masses. 
Gluino-squark associated production also leads to the 5~jets + $\MET$ missing topology; in this case all squarks are heavier than the gluino and decay via $\tilde q\to q\tilde g$, and the gluino then decays further to two jets and the $\tilde\chi^0_1$. This is the dominant missing topology for compressed gluino and neutralino masses in the bino-like LSP case, see the blue points in the left panel of Figure~\ref{fig:GluMissingTopo}. 
When compressing the gluino and LSP masses even further, such that the gluino decay is not visible any more, this gives jet + $\MET$ (dark green points), which is however a rather fine-tuned situation in the pMSSM and thus occurs much less often. 

Also noteworthy are the orange points, which denote an asymmetric 2~jets + $\MET$ topology 
with the two jets on one branch and nothing on the other branch. 
This can come from $\tilde\chi^0_1\tilde\chi^0_{i\not=1}$, $\tilde\chi^0_1\tilde\chi^\pm_{1,2}$ or
$\tilde\chi^0_1\tilde g$ associated production.
While EW-ino and $\tilde\chi^0_1\tilde g$ production can have comparable cross sections, the latter process is often disregarded. 
(The same topology can also arise from gluino-squark associated production when $\tilde g\to q\bar q\tilde\chi^0_1$ and the $\tilde q$ 
decay is ``invisible'' because of mass compression with the LSP.) 
Other topologies like $2b2t$+jet+$\MET$ (from $pp\to \tilde g\tilde q$, $\tilde q\to q\tilde g$, $\tilde g\to tb\chi^+_1$ in the higgsino-LSP case) 
or long cascades with $4b$+jet+$\MET$ or 3~jets+$2W$+$\MET$ also show up in Figure~\ref{fig:GluMissingTopo}, but are much less often 
the missing topology with highest cross section.
The corresponding diagrams, SUSY processes and labelling in \smodels\ notation can be found in Appendix~A. 

We note that all these missing topologies could be constrained from the 
${\rm jets}+\met$ searches, if the appropriate SMS interpretations were available. 
For instance, a limit of 40, 20, 10~fb on the 3~jets + $\MET$ missing topology cross section would exclude additional 4846, 5799, 6599 
(1377, 1948, 2637) points of the bino-like (higgsino-like) LSP dataset.  
We have explicitly checked  a couple of representative \smodels-allowed points with a high 3~jets + $\MET$ cross section and verified that
including the efficiencies for the relevant gluino-squark simplified model would indeed exclude these points. A specific example is provided
in Appendix~B.

\subsection{Third generation}

\begin{figure}[ht!]\centering
\includegraphics[width=0.5\textwidth]{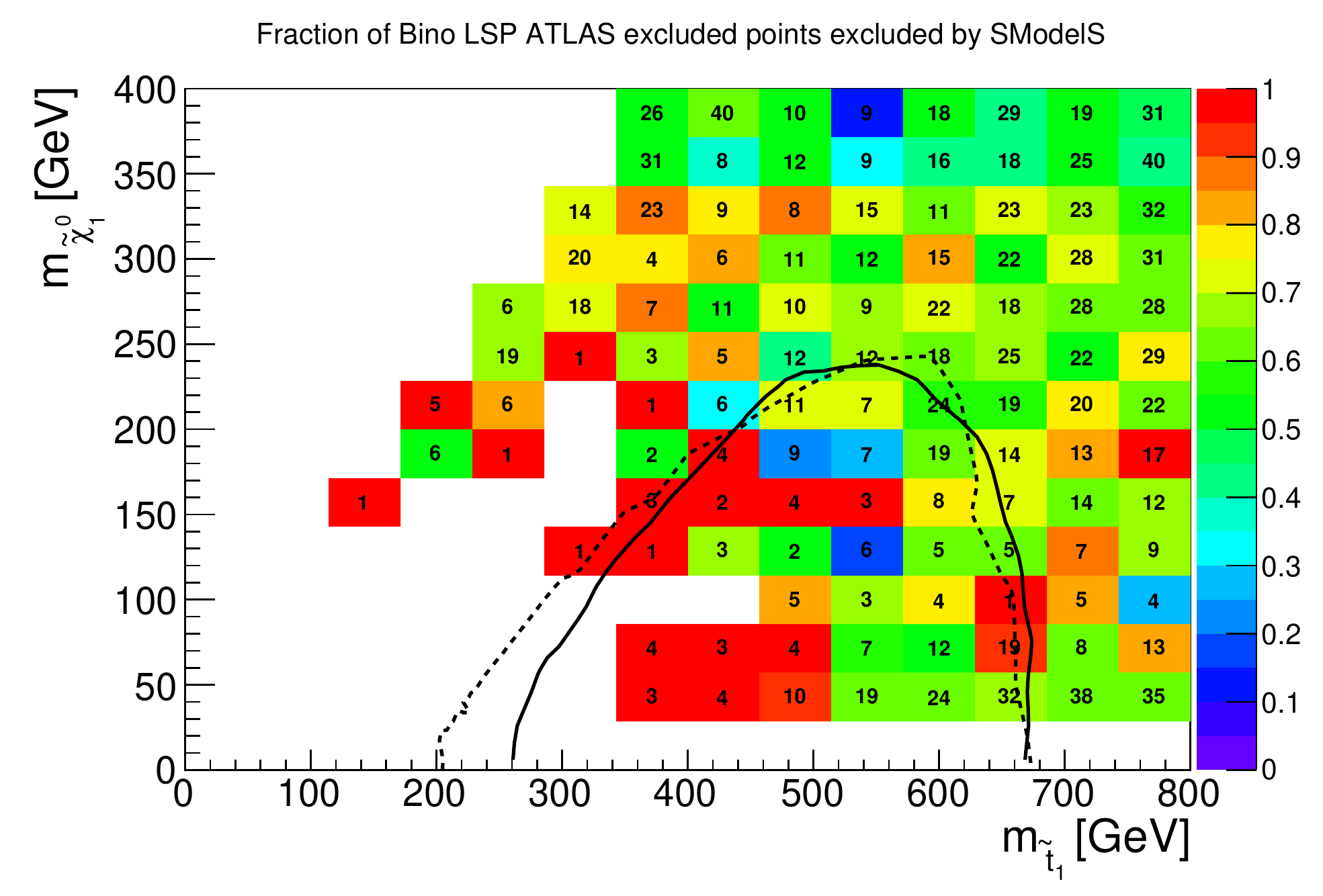}%
\includegraphics[width=0.5\textwidth]{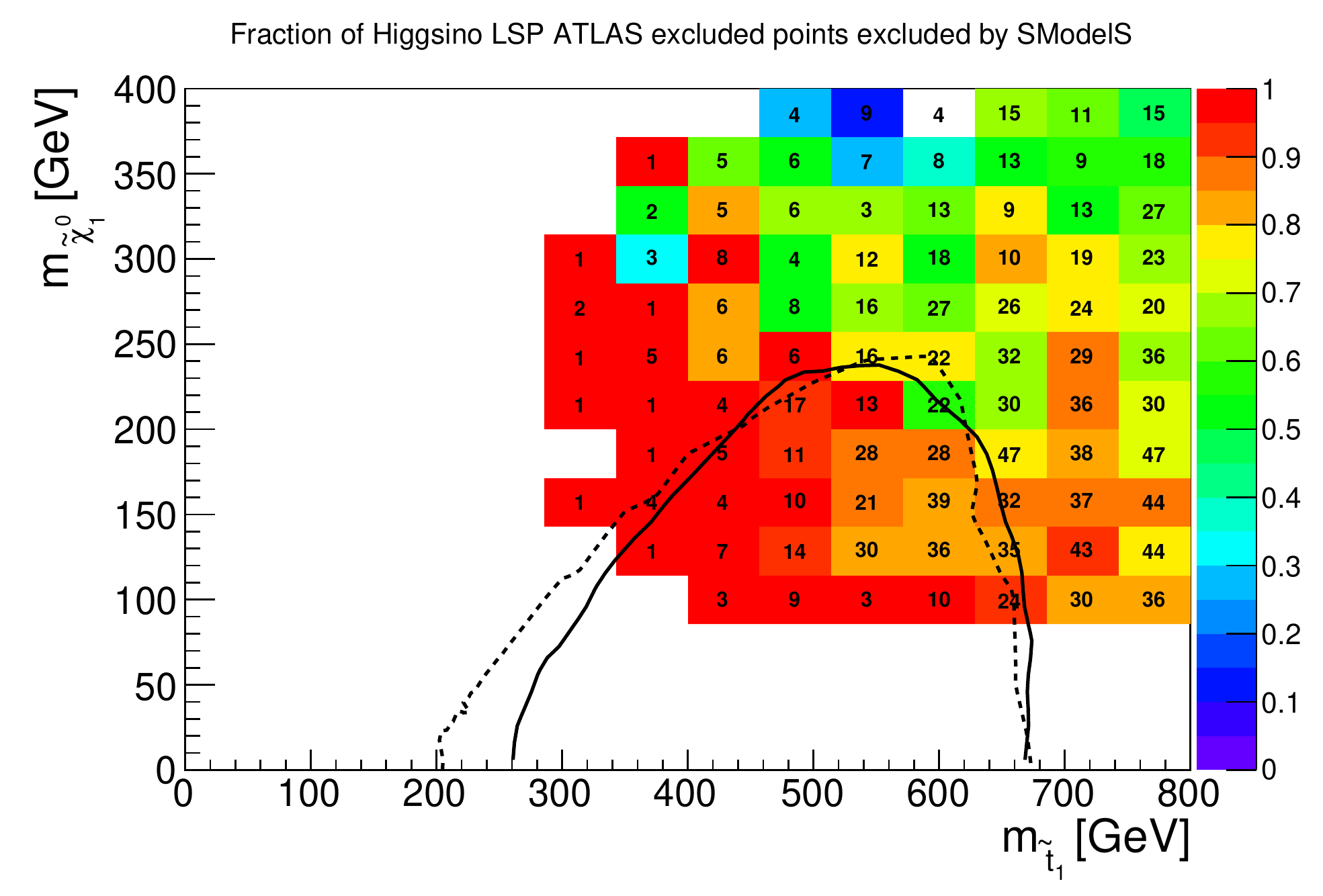}
\caption{Coverage in the stop vs.\ neutralino mass plane, for $\tilde t_1$ masses up to 800~GeV, 
for bino-like LSP scenarios (left) and higgsino-like LSP scenarios (right).
The color code indicates the fraction of points excluded by \smodels\ as compared to ATLAS, 
while the text gives the total number of points tested in each bin. 
For comparison, the black lines are the 95\%~CL exclusion curves for the $\tilde t_1\to t\tilde\chi^0_1$ simplified model 
from~\cite{Aad:2014bva} (0-lepton mode, full line) and \cite{Aad:2014kra} (1-lepton mode, dashed line). 
\label{fig:stop-n1-2d}}
\end{figure}

\begin{figure}[t!]\centering
\includegraphics[width=0.5\textwidth]{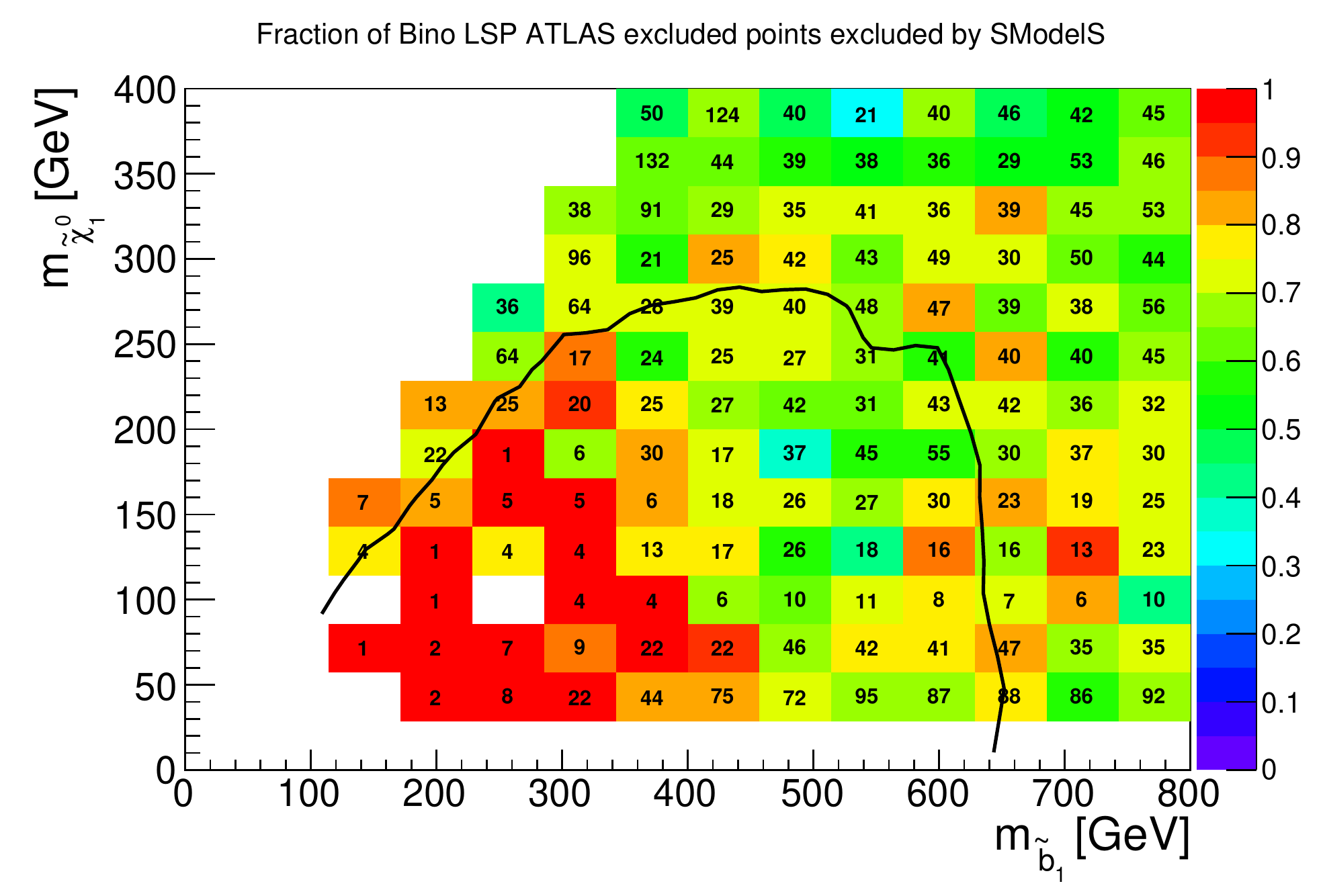}%
\includegraphics[width=0.5\textwidth]{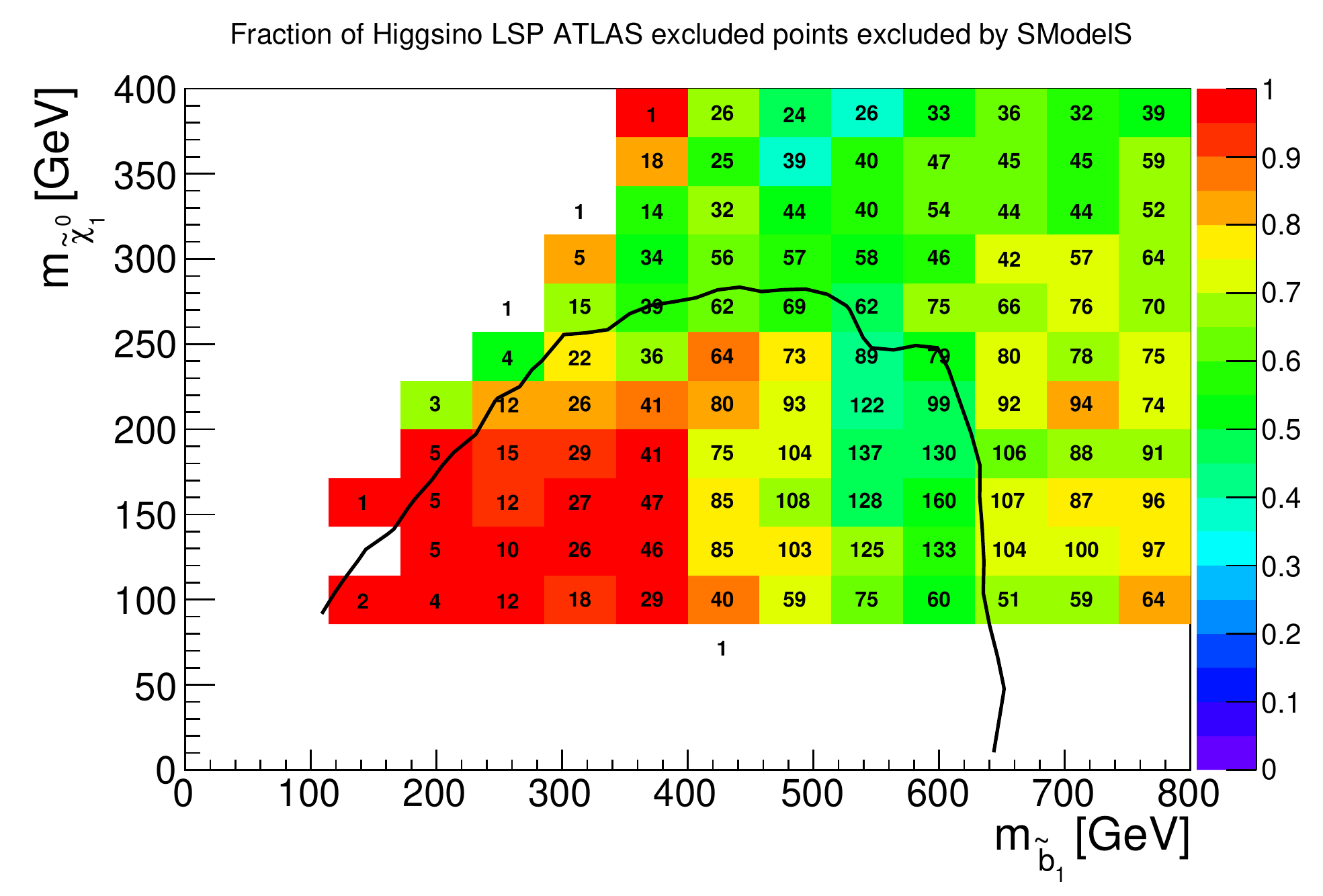}
\caption{Coverage in the sbottom vs.\ neutralino mass plane, for $\tilde b_1$ masses up to 800~GeV, 
for bino-like LSP scenarios (left) and higgsino-like LSP scenarios (right).
The color code indicates the fraction of points excluded by \smodels\ as compared to ATLAS, 
while the text gives the total number of points tested in each bin. 
The black line is the 95\%~CL exclusion line for the $\tilde b_1\to b\tilde\chi^0_1$ simplified model from~\cite{Aad:2013ija}. 
\label{fig:sbot-n1-2d}}
\end{figure}

Apart from gluinos and squarks, which may be regarded as the primary (and easiest) targets of the SUSY searches, 
searches for stops and sbottoms are of particular interest.  The coverage obtained by \smodels\ in the 
stop vs.\ neutralino  and sbottom vs.\ neutralino mass planes is shown in Figures~\ref{fig:stop-n1-2d} and \ref{fig:sbot-n1-2d}.
We also show the official exclusion curves for the $\tilde t_1\to t\tilde\chi^0_1$ and $\tilde b_1\to b\tilde\chi^0_1$ simplified models  
from~\cite{Aad:2014bva,Aad:2014kra,Aad:2013ija}, to help identify the region expected to be excluded by stop or sbottom production only. 

\begin{figure}[t!]\centering
\includegraphics[width=0.5\textwidth]{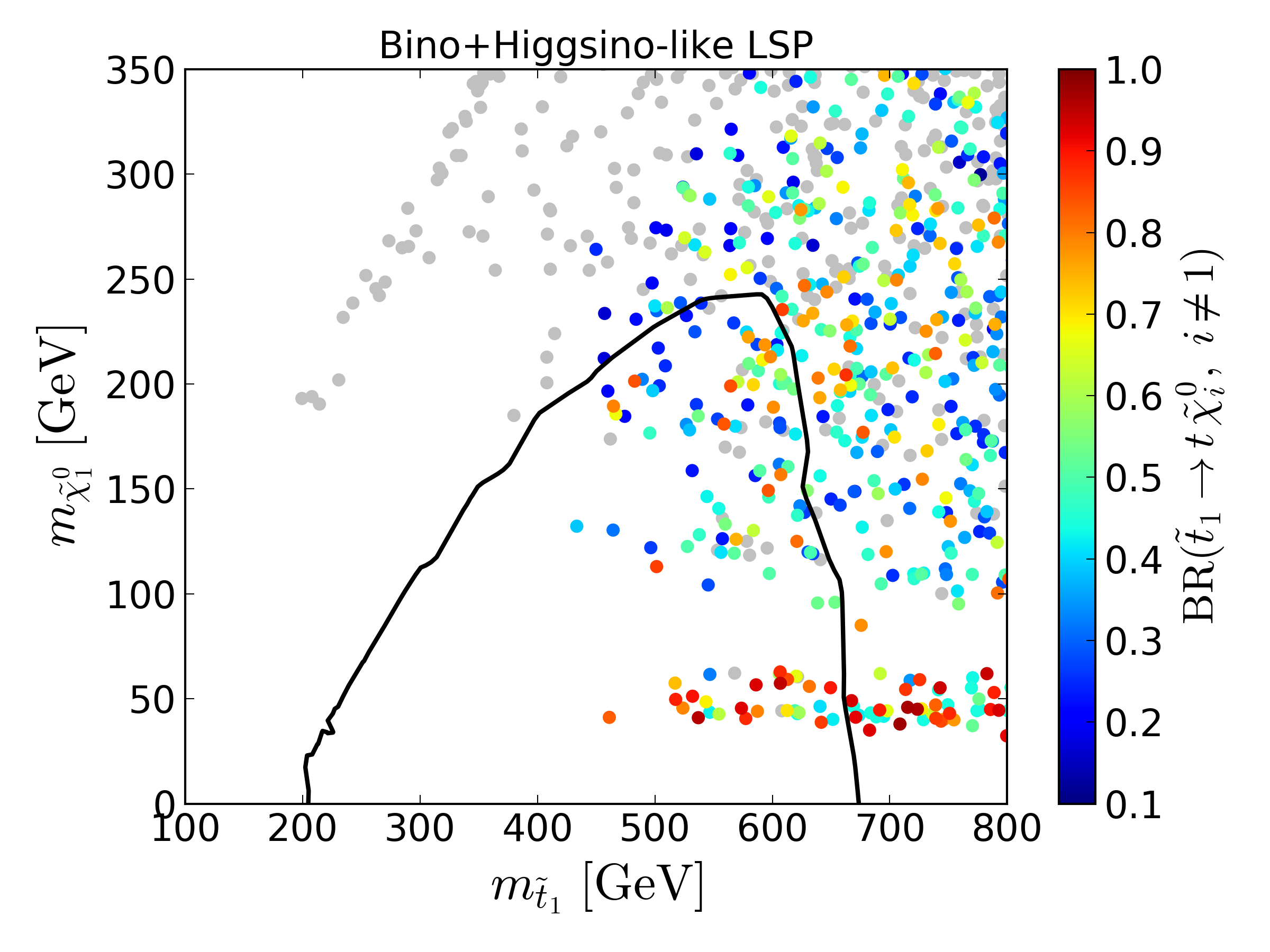}%
\includegraphics[width=0.5\textwidth]{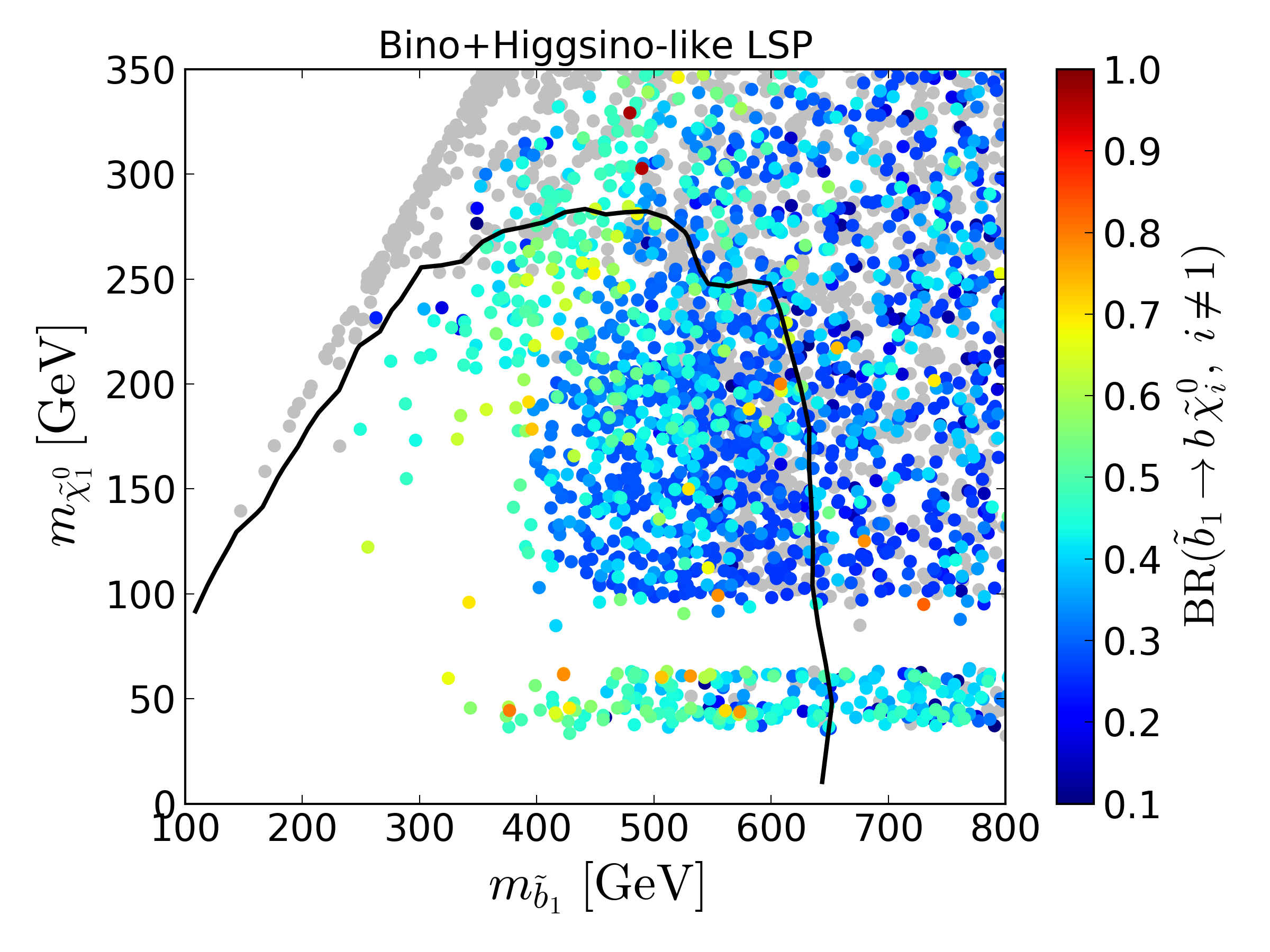}
\caption{Branching ratios of stop (left) and sbottom (right) decays into heavier neutralino mass eigenstates for \smodels-allowed points, leading to signatures for which no SMS results are currently available ($m_{\tilde\chi^0_{i\not=1}}\!\!-m_{\tilde\chi^0_1}\ge5$~GeV). Here, bino- and higgsino-like LSP scenarios are  combined. Grey points have ${\rm{BR}}<10\%$ for the decays considered. The black lines are the 95\%~CL exclusion lines for the $\tilde t_1\to t\tilde\chi^0_1$ simplified model from~\cite{Aad:2014kra} (left) and the $\tilde b_1\to b\tilde\chi^0_1$ simplified model from~\cite{Aad:2013ija} (right). See text for details. 
\label{fig:stop-sbot-branchings}}
\end{figure}

For stops, we observe an excellent coverage in the higgsino-LSP case when compared to
the official exclusion curves. (A slightly stronger exclusion is obtained by the combination of the 0-lepton and 1-lepton analyses~\cite{Aad:2015pfx}, but no UL maps are available for the combination.)
Contrary to the gluino case, the stop exclusion is not driven by EM results but by the UL maps for $t\bar t+\met$ and  $b\bar b+\met$ final states (mostly because not so many different EMs are available for stops and sbottoms).  
Points outside the naive SMS exclusion line are excluded by other searches or because of light sbottoms which also contribute to the signal.  
In the bino-LSP case, on the other hand, light stops in the 500--650~GeV mass range often escape exclusion by SMS results. This is mostly  
because they share out their branching ratios in $\tilde t_1\to t\tilde\chi^0_2\to tZ\tilde\chi^0_1$ and $\tilde t_1\to b\tilde\chi^+_1\to bW\tilde\chi^0_1$ 
cascade decays. While we do have EMs for a so-called T6bbWW simplified model, \ie\ a $2b2W+\met$ final state originating from both stops decaying via an intermediate chargino, the equivalent topologies for one or both stops decaying via an intermediate neutralino (\eg, $tbWZ+\met$ and $2t2Z+\met$ final states) are missing. Including EMs for these topologies for a variety of intermediate $\tilde\chi^0_2$ and $\tilde\chi^\pm_1$ masses would certainly allow us to get closer to the ATLAS exclusion.\footnote{Note that for cascade decays via an intermediate sparticle, it is important to have several mass planes in order to be able to interpolate in all dimensions of the SMS; see also Appendix~C of \cite{Ambrogi:2017neo}.} 
Notice, however, that for light stops we are dealing with small numbers of points in each bin, so large fluctuations in the coverage are easily possible. 
The importance of $\tilde t_1\to t\tilde\chi^0_{i\not=1}$ decays, followed by visible $\tilde\chi^0_{i\not=1}$ decays, for \smodels-allowed points is illustrated in the left plot in Figure~\ref{fig:stop-sbot-branchings}.

Turning to sbottoms, we see that the coverage is quite good for 
$m_{\tilde b_1}\lesssim 450$~GeV and $m_{\tilde\chi^0_1}\lesssim 250$~GeV.
For these mass ranges, $\tilde b_1\to b\tilde\chi^0_1$ (and/or $\tilde b_1\to t\tilde\chi^-_1$ in the higgsino-LSP case) decays dominate.  
Once a larger variety of decay channels becomes relevant, the exclusion drops to about 50\% of that of ATLAS. 
While results for $\tilde b_1\to t\tilde\chi^-_1\to tW\tilde\chi^0_1$ are available from ATLAS~\cite{ATLAS-CONF-2013-007} and CMS~\cite{Chatrchyan:1631468},\footnote{We appreciate the fact that these are given for 3 different chargino masses.} 
these are ULs for a same-sign lepton signature assuming both sbottoms decay via a chargino; they have a reach in sbottom mass of at most 500--550~GeV.  
It would be useful to have in addition simplified model results for $\tilde b_1\to b\tilde\chi^0_{i\not=1}\to bZ\tilde\chi^0_1$ or $bh\tilde\chi^0_1$,  best in the form  of EMs for symmetric and asymmetric decay branches. The importance of these decay modes for \smodels-allowed points is illustrated in the right plot in Figure~\ref{fig:stop-sbot-branchings}. 

It is relevant to stress that the branching ratios shown in Figure~\ref{fig:stop-sbot-branchings} only
consider {\it visible} decays.  
In particular the higgsino-like LSP dataset contains many points where sbottom branching ratios are shared out in $\tilde b \to b\tilde\chi^0_{i\not=1}$ and $t\tilde\chi^-$ decays (contributing to the reduced coverage for $m_{\tilde b_1}\gtrsim 500$~GeV seen in Figure~\ref{fig:sbot-n1-2d}) but the subsequent EW-ino decays are invisible because of mass compression.
This leads to the patch of grey points just below the exclusion curve in the right plot of Figure~\ref{fig:stop-sbot-branchings}. Regardless of this, the conclusion from Figure~\ref{fig:stop-sbot-branchings} is that EM results for stops and sbottoms decaying through an intermediate particle (leading to final states with additional $W$, $Z$ or $h$ bosons) would be highly desirable.

\subsection{EW production}

\begin{figure}[t!]\centering
\includegraphics[width=0.5\textwidth]{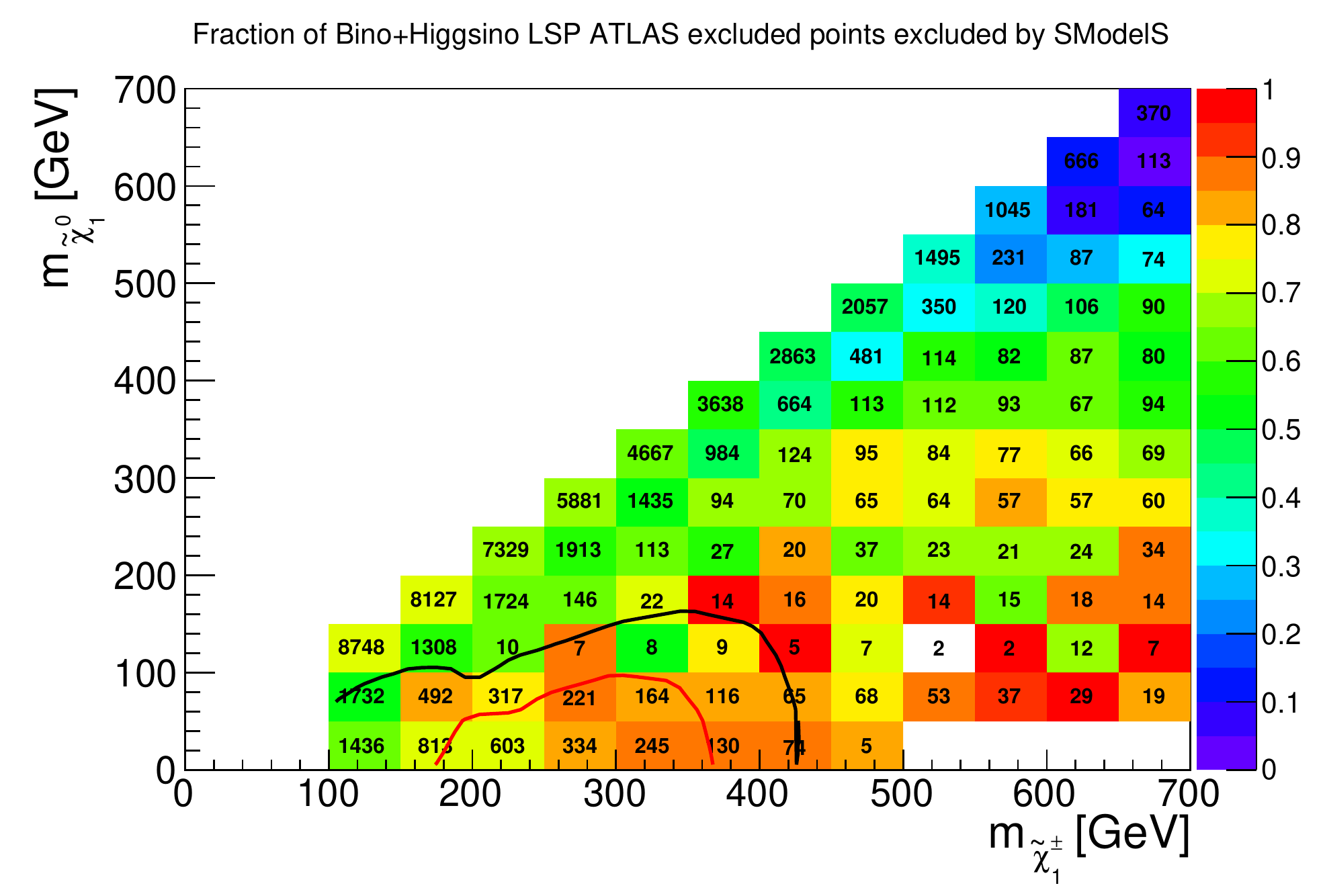}%
\includegraphics[width=0.5\textwidth]{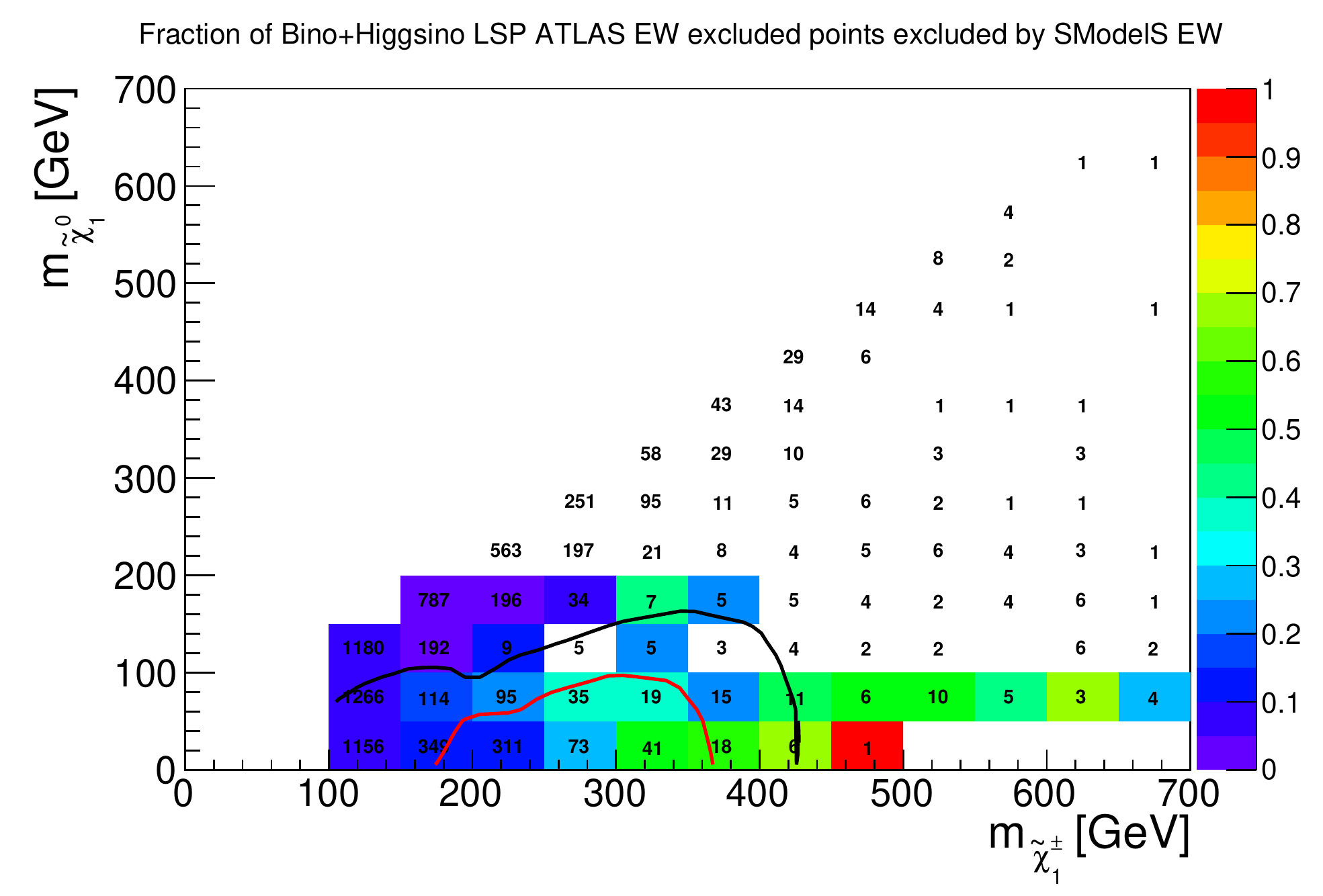}
\caption{Coverage in the chargino vs.\ neutralino mass plane, for $\tilde\chi^\pm_1$ masses up to 700~GeV. 
Here, bino-like and higgsino-like LSP scenarios have been combined to increase the number of points.
The plot on the left considers all analyses, the plot on the right only EW analyses. 
The color code indicates the fraction of points excluded by \smodels\ as compared to ATLAS, 
while the text gives the total number of points tested in each bin. 
For comparison, the exclusion line from the \ThreeLepton\ analysis~\cite{Aad:2014nua} is shown in red 
and from the combination paper~\cite{Aad:2015eda} is drawn in black. 
\label{fig:charg-neut-2d-all}}
\end{figure}

It is also interesting to study how well EW production is covered by simplified models. 
To this end, we first show in Figure~\ref{fig:charg-neut-2d-all} the coverage in the chargino vs.\ neutralino LSP mass plane.
Here, the bino-like and higgsino-like LSP scenarios have been combined to increase the number of points. 
In the plot on the left, light charginos seem to be reasonably well constrained. However, this does not come 
from searches looking specifically for EW production, as is apparent from the plot on the right.  
The fact that the coverage does not follow the SMS exclusion curve is no surprise, as the latter was 
obtained for the best-case scenario of pure wino production. However, from the color code we see that the 
constraining power of EW searches is very poorly reproduced by SMS results. One of the reasons is that 
the SMS results typically assume strictly mass-degenerate $\tilde\chi^\pm_1$ and $\tilde\chi^0_2$, a condition which is 
rarely satisfied in the pMSSM. Moreover, BR($\tilde\chi^0_{i\not=1}\to h\tilde\chi^0_1$) is often sizeable, 
which further reduces the coverage. (The SMS limit in the $Wh+\met$ final state is effective only for very light LSP below 40 GeV and cannot be combined with the limit on the $WZ+\met$ final state.)
Finally, the 3 or 4 lepton searches in ATLAS do not have a jet veto; therefore in the ATLAS pMSSM study strong production may also feed into the EW exclusion, which is not the case in \smodels\ for lack of the corresponding SMS results.

\begin{figure}[t!]\centering
\includegraphics[width=0.5\textwidth]{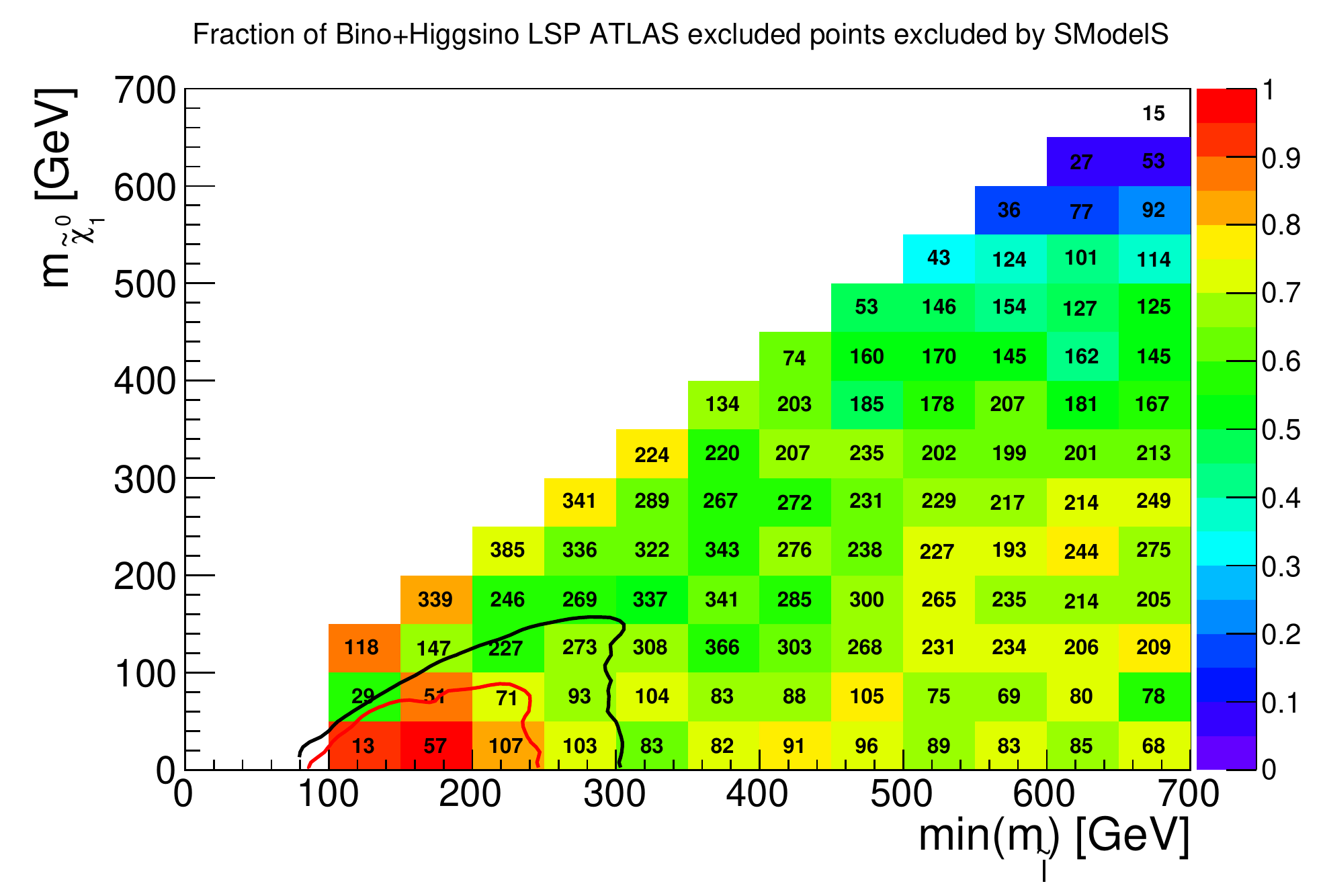}%
\includegraphics[width=0.5\textwidth]{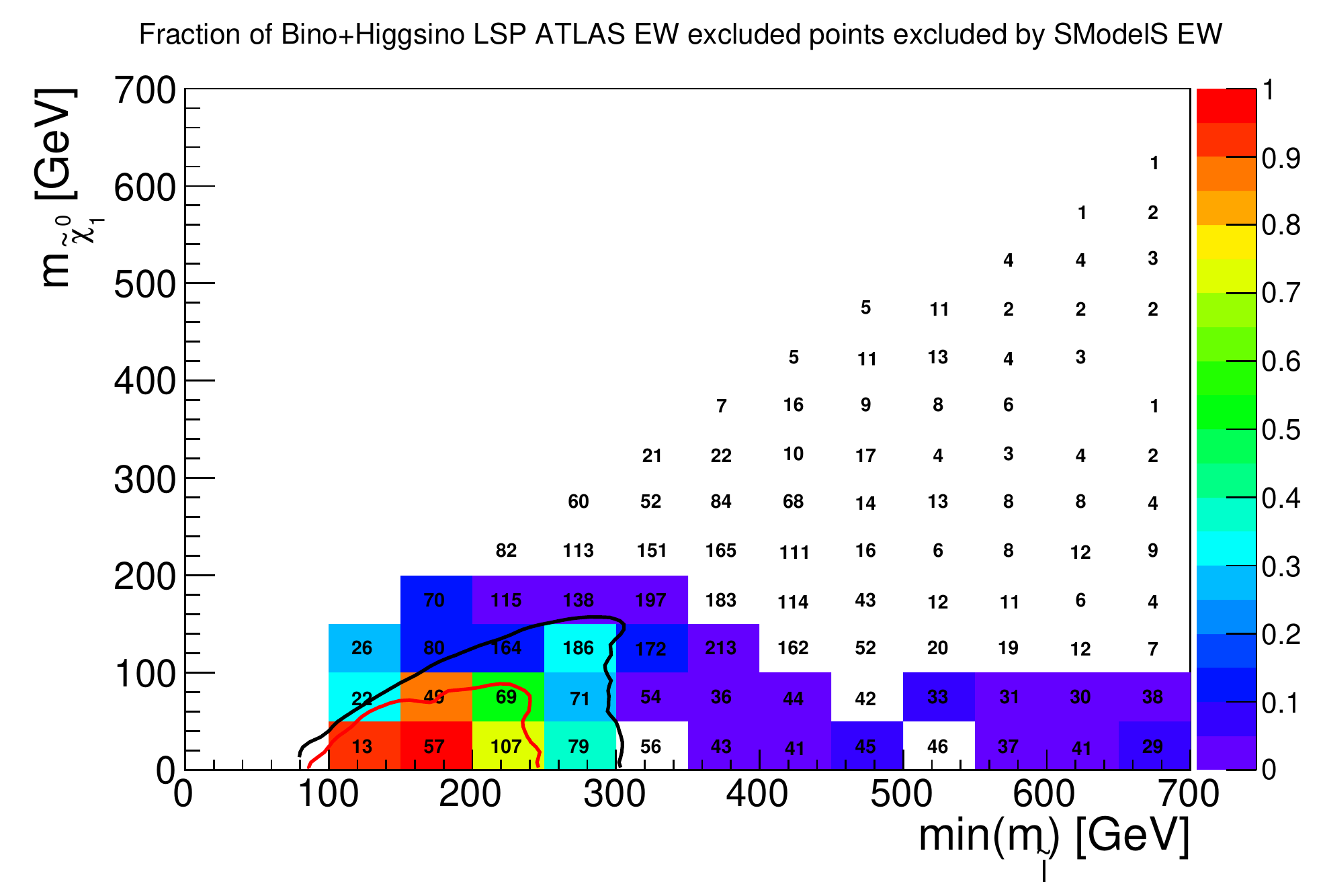}
\caption{Coverage in the plane of lightest slepton (first two generations) vs.\ LSP mass, for $\tilde l$ masses up to 700~GeV. 
Here, bino-like and higgsino-like LSP scenarios have been combined to increase the number of points.
The plot on the left considers all analyses, the plot on the right only EW analyses. 
The color code indicates the fraction of points excluded by \smodels\ as compared to ATLAS, 
while the text gives the total number of points tested in each bin. 
The exclusion lines for $\tilde l_R$ (red) and $\tilde l_L$ (black) are also shown for comparison. 
\label{fig:slep-neut-2d}}
\end{figure}

In Figure~\ref{fig:slep-neut-2d} we show the same kind of plots for sleptons. 
Here, the coverage is quite good and reproduces reasonably well the SMS exclusion line for right sleptons. 
The exclusion line for left sleptons is naturally matched less well, because pMSSM points contain a mix of light left and right sleptons. 
Finally, a small fraction of points with $min(m_{\tilde l})=250$--$300$~GeV and light LSP escape exclusion in \smodels\ because the sleptons 
partly undergo cascade decays via heavier EW-inos.  Even if the direct decay into the LSP still dominates, the reduction in BR 
can be enough to result in $r<1$. 

\begin{figure}[t!]\centering
\includegraphics[width=0.5\textwidth]{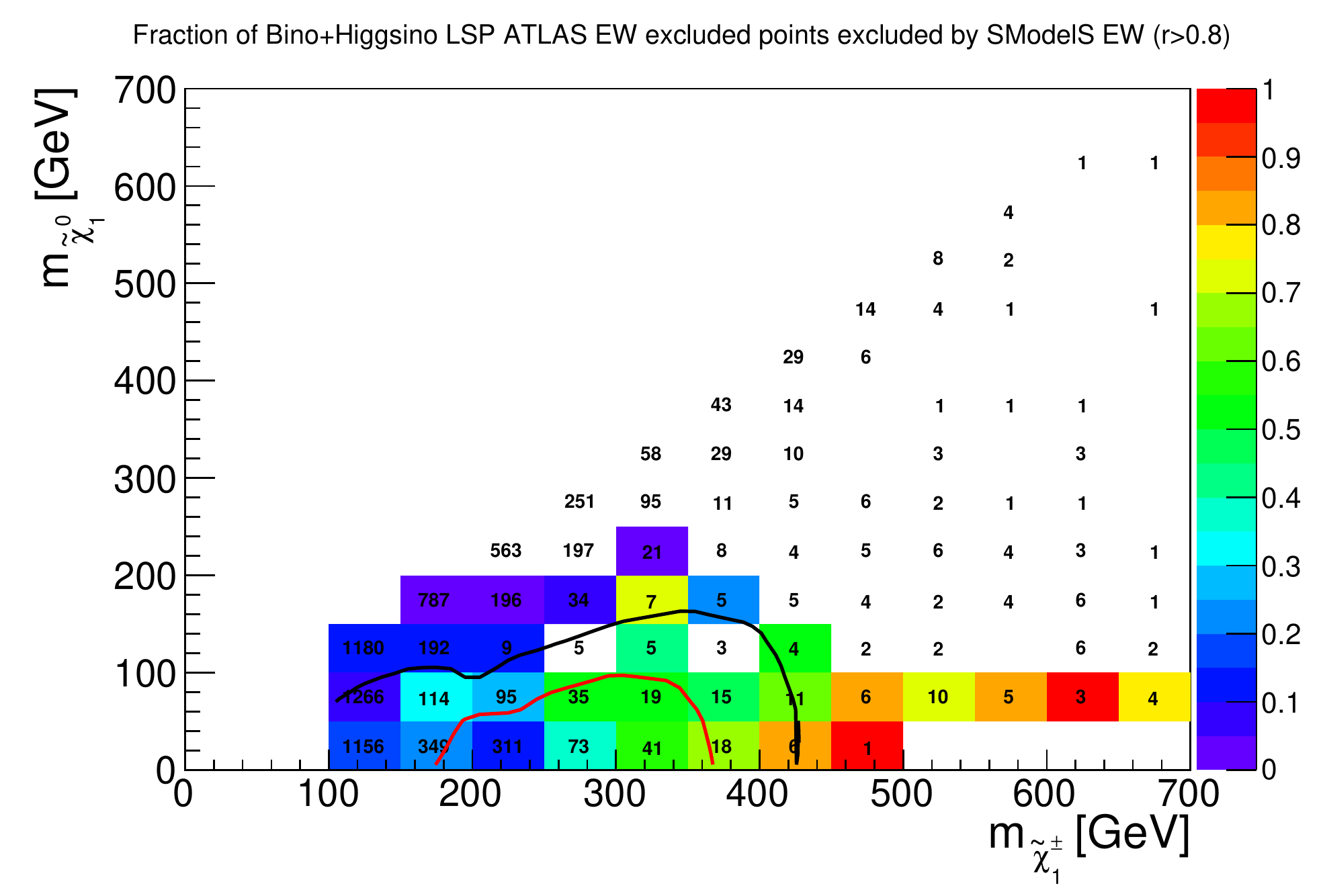}%
\includegraphics[width=0.5\textwidth]{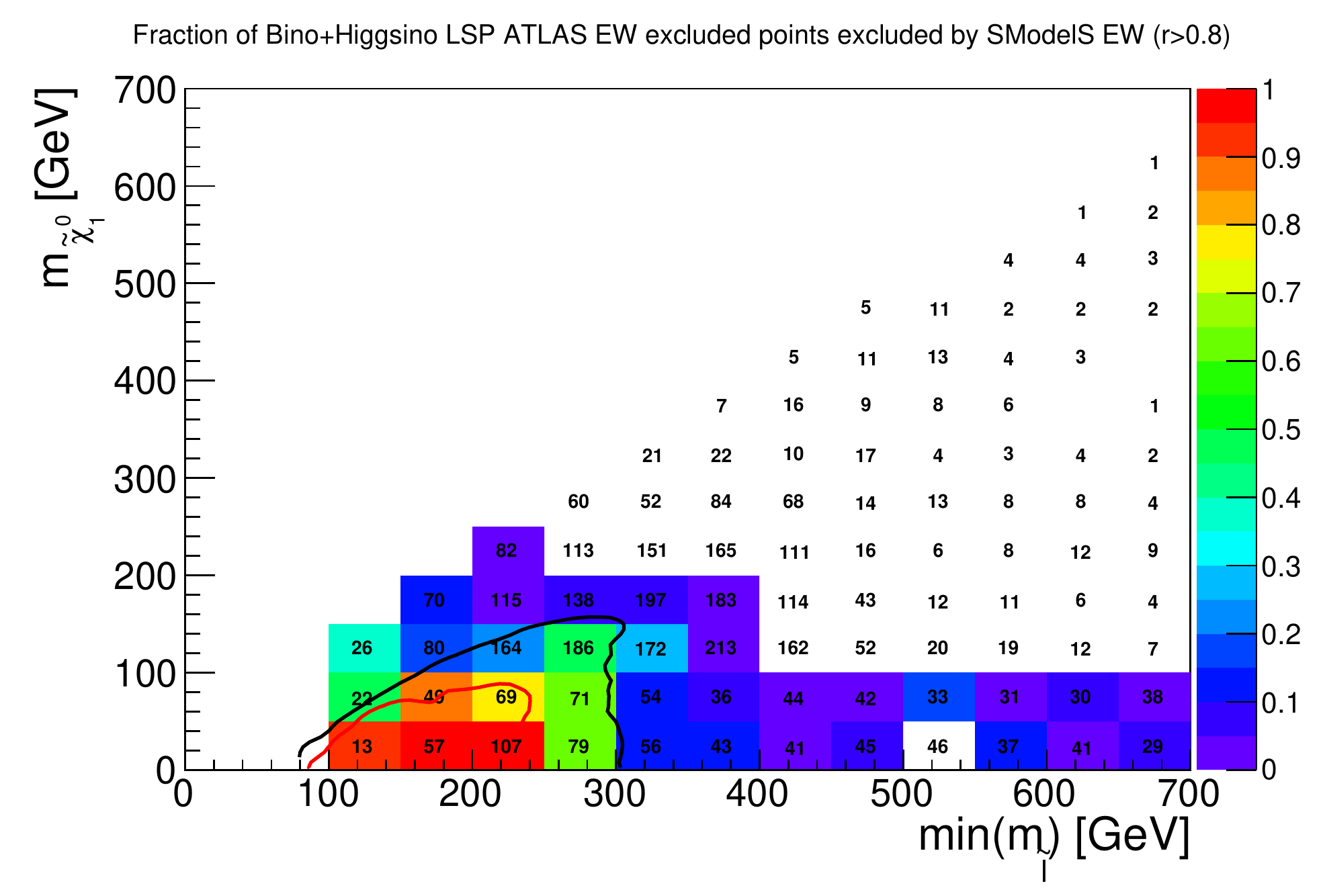}
\caption{Coverage of EW-inos and sleptons by EW analyses analogous to the right plots of Figures~\ref{fig:charg-neut-2d-all} 
and \ref{fig:slep-neut-2d} but considering points with $r_{\rm max}>0.8$ (instead of $r_{\rm max}>1$) as excluded. 
\label{fig:EW-2d-rescaled}}
\end{figure}

Last but not least we recall that EW cross sections are computed at leading order in \smodels. 
Radiative corrections typically increase these cross sections by about 20\%, which slightly improves the coverage of the EW sector 
but does not change the overall picture. This is illustrated in Figure~\ref{fig:EW-2d-rescaled}, which shows the coverage of EW-inos and sleptons by EW analyses when rescaling the relevant $r$ values by 20\%.  

\section{Conclusions}\label{sec:conclusions}

We studied to which extent the SUSY search results published by ATLAS and CMS in the context of SMS constraints
actually cover the more realistic scenarios of a full model, concretely the phenomenological MSSM.
To this end we analysed the exclusion obtained with \smodels~\cite{Ambrogi:2017neo,Kraml:2013mwa} with respect to the 
ATLAS pMSSM study~\cite{Aad:2015baa}. From about 84K pMSSM points excluded by ATLAS, the 8~TeV results in \smodelsv\ 
exclude about 50K points. 
Efficiency maps proved to be important for constraining scenarios with a variety of 
production and/or decay modes, because they allow to combine different contributions to the same signal region. 
Nonetheless, despite the plethora of SMS results available, about 40\% of the points excluded by ATLAS are not excluded by \smodels.  
These ``escaping'' points include gluinos as light as about 500~GeV, but also light stops/sbottoms or EW-inos with reasonably large cross sections. 
We analysed the reasons for this limited coverage and how it might be improved. 

Concretely, we found that a large part of the unconstrained cross section goes into simple but asymmetric topologies, 
either because pair-produced sparticles have two or more relevant decay modes, or because of associated production 
of two different sparticles. A particularly important case, for which no SMS results are currently available, is a 3-jet topology 
stemming from gluino--squark associated production with non-degenerate squarks: $pp\to \tilde g\tilde q$ followed by 
$\tilde g\to q\tilde q\to q\bar q \tilde\chi^0_1$ and $\tilde q\to q\tilde\chi^0_1$ when one of the squarks is lighter than the gluino, 
or $\tilde g\to g\tilde\chi^0_1$ and $\tilde q\to q\tilde g\to qg \tilde\chi^0_1$ otherwise. For one third of the bino-like LSP points which 
are excluded by ATLAS but not by \smodels, this topology has a cross section $>20$~fb. 

For the case that the produced SUSY particles share out their branching ratios over several different decay modes, 
which need to be combined to obtain a good limit, we highlighted the example of stop and sbottom decays via heavier EW-inos, 
which in turn decay visibly into the LSP.   
While SMS results for stop-pair production with both stops decaying via an intermediate chargino exist, analogous results considering also 
$\tilde t_1\to t\tilde\chi^0_2\to tZ\tilde\chi^0_1$, $\tilde b_1\to b\tilde\chi^0_2\to bZ\tilde\chi^0_1$ or $\tilde b_1\to t\tilde\chi^-_1\to tW\tilde\chi^0_1$ 
decays are missing. Efficiency maps for these cases would be highly desirable to improve the coverage of the third generation. 
 
Regarding the EW SUSY sector, the coverage of light sleptons by SMS results is quite good. For EW-inos, however, the situation is less satisfying. 
This might be improved if EMs were available for the EW-ino searches in multi-lepton channels instead of only UL-type results. Moreover, for multi-lepton 
searches without jet veto, EM results applicable also to EW-inos stemming from strong production would be interesting. 

The coverage in \smodels\ may also be limited when the initially produced SUSY particles undergo a series of cascade decays  
leading to long decay chains with more than one intermediate sparticles. This situation is difficult to cover by simplified models, 
since it involves a large number of free parameters. 
Interestingly, we find that only a small fraction of the points which escape exclusion by \smodels\ fall into this class.  
In this view it is much more useful to improve the constraining power of simple SMS (with few parameters) by providing, \eg, 
additional efficiency maps and sufficient mass-vs-mass planes for a reliable interpolation in all mass dimensions, 
than to present results for more complicated topologies. 
Although complicated topologies (decay chains with more than 3 mass parameters) have been considered by the experimental collaborations, 
these results always assume very specific mass relations to limit the number of free parameters and hence cannot be used for generic scenarios. 

Overall, the SMS approach provides a powerful means to quickly test the predictions of new physics models against the constraints from a large variety of experimental searches. 
However, not excluded by SMS results does not automatically mean allowed by all LHC searches; it is advisable to further test ``surviving" points with Monte Carlo event simulation, if they have sizeable cross sections. Implementations of ATLAS and CMS analyses in public recasting tools like 
{\sc CheckMATE}~\cite{Drees:2013wra,Dercks:2016npn}, {\sc MadAnalysis}\,5~\cite{Conte:2014zja,Dumont:2014tja}, {\sc Rivet}~\cite{Buckley:2010ar} (v2.5 onwards) and {\sc GAMBIT}'s ColliderBit~\cite{Athron:2017ard,Balazs:2017moi} can be used to this end.  
Finally, these tools may also be used to produce additional SMS results beyond those provided by the experimental collaborations.

\section*{Acknowledgements} 

We thank the ATLAS and CMS SUSY groups for providing a plethora of SMS cross section upper limits and efficiency
maps in digital format. Moreover, we are particularly grateful for the detailed material on the ATLAS pMSSM study 
available on HepDATA, which made the analysis presented in this paper possible. 
 
This research was supported in part by the ANR project DMASTROLHC grant no.\ ANR-12-BS05-0006,  
the IN2P3 project ``Th\'eorie -- LHCiTools'' and the CNRS-FAPESP collaboration PRC275431. 
F.A.\ is supported by the Austrian FWF, project P26896-N27, 
Su.K.\ by the ``New Frontiers'' program of the Austrian Academy of
Sciences, U.L.\ by  the ``Investissements d'avenir, Labex ENIGMASS'', and 
A.L.\  by the S\~ao Paulo Research Foundation (FAPESP), projects 2015/20570-1 and 2016/50338-6.  

\clearpage

\begin{appendix}

\section{Diagrams and processes for missing topologies}
\label{sec:missingtopotable}
 
Here we show the explicit diagrams, SUSY processes and labelling in \smodels\ notation for the missing 
topologies of Figure~\ref{fig:GluMissingTopo}.  \\[4mm]

\noindent
\begin{tabular}{ m{5cm} m{6cm} m{1cm} m{5cm}} \hline
\mbox{Short label in Fig.~\ref{fig:GluMissingTopo}}, \mbox{\smodels\ notation} & main SUSY process(es) \hfill & & graph \\ \hline
  \mbox{3~jets + $\MET$}, \mbox{[[[jet]],\,[[jet],[jet]]]}  & 
  \mbox{$\tilde g\tilde q$, $\tilde g\to q\tilde q$, $\tilde q\to q\tilde\chi^0_1$; or} \hfill 
  \mbox{$\tilde g\tilde q$, $\tilde q\to q\tilde g$, $\tilde g\to g\tilde\chi^0_1$} & &
  \includegraphics[width=3cm]{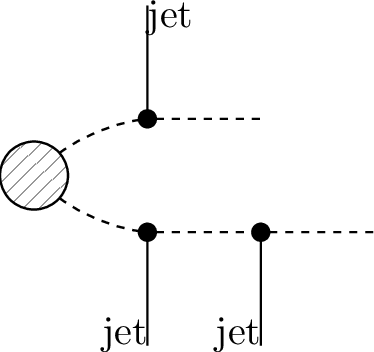} \\[5mm] 
  \mbox{5~jets + $\MET$}, \mbox{[[[jet],[jet,jet]],\,[[jet,jet]]]}  & 
  $\tilde g\tilde q$, $\tilde q\to q\tilde g$, $\tilde g\to q\bar q\tilde\chi^0_1$ & &
  \includegraphics[width=3cm]{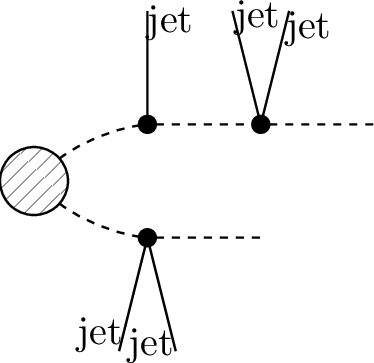} \\[5mm] 
  \mbox{$2b2t$ + jet + $\MET$}, \mbox{[[[b,t]],[[jet],[b,t]]]}  & 
  $\tilde g\tilde q$, $\tilde q\to q\tilde g$, $\tilde g\to bt\tilde\chi^0_1$ & &
  \includegraphics[width=3cm]{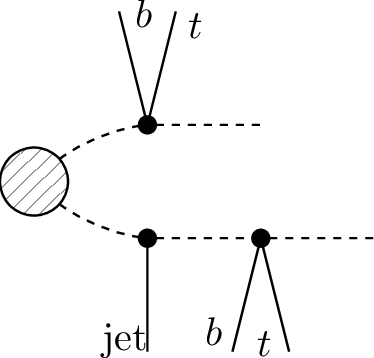} \\[5mm] 
  \mbox{jet + $\MET$}, \hphantom{bla bla bla} \mbox{[[[],[[jet]]]}  & 
  $\tilde g\tilde q$, $\tilde q\to q\tilde g$, ($\tilde g\to q\bar q \chi^0_1$ or $g \chi^0_1$ being invisible) & &
  \includegraphics[width=2.6cm]{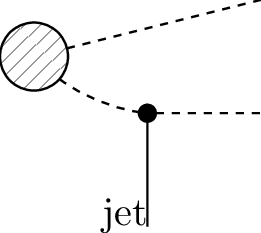} \\[5mm] 
  \mbox{2~jets + $\MET$}, \hphantom{bla bla bla} \mbox{[[[],[[jet,jet]]]}  & 
  $\tilde\chi^0_1\tilde\chi^0_{i\not=1}$, $\tilde\chi^0_1\tilde\chi^\pm_{1,2}$, 
  $\tilde\chi^0_1\tilde g$ production followed by decay to $q\bar q\tilde\chi^0_1$; or 
  \mbox{$\tilde g\tilde q$ production} for compressed squark and LSP & &
  \includegraphics[width=2.6cm]{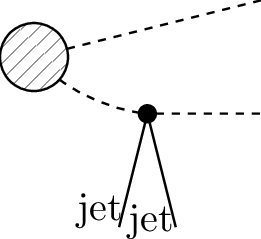} \\[5mm] 
\end{tabular}

\noindent
\begin{tabular}{ m{5cm} m{6cm} m{1cm} m{5cm}} 
  \mbox{$4b$ + jet + $\MET$}, \mbox{[[[b],[b]],\,[[jet],[b],[b]]]}  & 
  $\tilde g\tilde q$, $\tilde q\to q\tilde g$, $\tilde g\to b \tilde b \to b\bar b\tilde\chi^0_1$ & &
  \includegraphics[width=3cm]{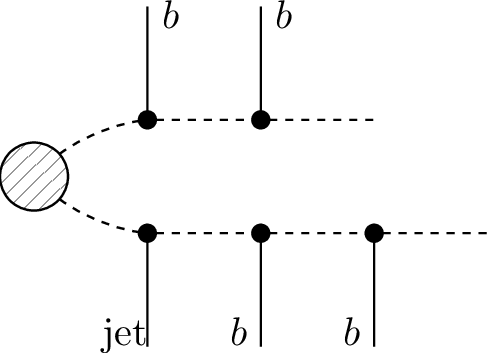} \\[5mm] 
  \mbox{4~jets + $\MET$}, \mbox{[[[jet]],\,[[jet],[jet,jet]]]}  & 
  $\tilde g\tilde q$, $\tilde q\to q\tilde g$, $\tilde g\to q\bar q \tilde\chi^0_1$ or $g\tilde\chi^0_1$ with comparable BR's; or 
$\tilde q \tilde q$, $\tilde q\to q \tilde\chi^0_1$ or $\tilde q\to q \tilde\chi^0_i / \tilde\chi^{\pm}_j, i \neq 1$ with comparable BR's
 & &
  \includegraphics[width=3cm]{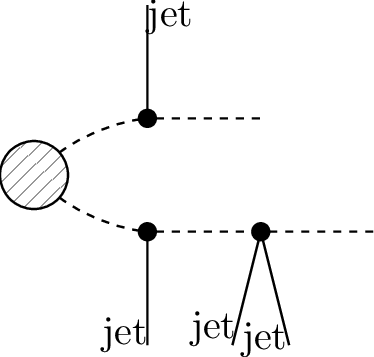} \\[5mm] 
  \mbox{3~jets + $2W$ + $\MET$}, \mbox{[[[jet],[W]],\,[[jet],[jet],[W]]]}  & 
  $\tilde g\tilde q$, $\tilde g\to q\tilde q$, $\tilde q\to q'\tilde\chi^\pm_1 \to q'W\tilde\chi^0_1$ & &
  \includegraphics[width=3cm]{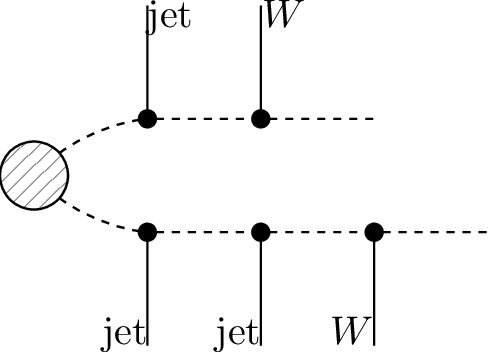} \\[5mm] 
\end{tabular}

\section{Example for the impact of a 3~jets + $\MET$ simplified model}
\label{sec:missingtopoexample}

In order to illustrate the importance of asymmetric topologies, we analyse here in more detail
one of the ATLAS-excluded points with a light gluino which has not been excluded by the SMS results.
The pMSSM point we consider is no.~192342466 of the bino-LSP dataset; it has light gluinos and a highly split spectrum of squarks  
with light $\tilde q_L^{}$. Concretely,
\begin{equation*}
   m_{\tilde{\chi}_1^0} = 666,~m_{\tilde g} = 712,~m_{\tilde u_L} = 758,~m_{\tilde d_L} = m_{\tilde s_L} = 761,~m_{\tilde d_R} = 1343,~m_{\tilde u_R} = 3968,  
\end{equation*}
where all values are in GeV. 
Stops and sbottoms are heavy with $m_{\tilde t_1,\tilde b_1}\approx 1.4$~TeV and  $m_{\tilde t_2,\tilde b_2}\approx 3.3$~TeV. 
The wino- and higgsino-like EW-inos have masses around $3.5$~TeV. 

In the following we only consider production of gluinos and the squarks 
$\tilde u_L,~\tilde d_L,~\tilde s_L,~\tilde d_R$, since this corresponds
to $\simeq 95\%$ of the total SUSY cross section for this point.
For simplicity we will refer to the associated and pair
production of these squarks as $\tilde g \tilde q$ and $\tilde q \tilde q$. 
The NLO+NLL cross section for gluino-pair production is 322~fb, while the $\tilde g\tilde q$ production cross section is 762~fb. 
The dominant gluino and squark decays are $\tilde g\to g + \tilde \chi^0_1$ (88\% BR) and $\tilde q^{}\to q\tilde g$ (99\% BR). 
As a result, a large fraction of the signal goes to the 3~jets + $\MET$ final state discussed as missing topology in Section~3.1. 

According to the ATLAS pMSSM study, this point is excluded by the \ZeroLepton\ search \cite{Aad:2014wea} (ATLAS-SUSY-2013-02). 
This is also the analysis which gives the highest $r$ value in \smodels, 
namely $r = 0.36$ for the $\tilde g \tilde g \to$ 2~jets + $\MET$ topology.
Hence this point is clearly not excluded by the SMS results.
In oder to investigate how specific topologies contribute to the 
total signal yield and to the exclusion of this point, 
we used the {\sc CheckMATE\,2} (v2.0.14)  implementation of this analysis along with Pythia\,8.230 and Delphes\,3.4.1 for event generation and detector simulation.
We generated signal events at leading order for associated and pair production of $\tilde g$, $\tilde u_L$, $\tilde d_L$, $\tilde s_L$ and $\tilde d_R$ and then rescaled
the cross sections using the K-factors computed with {\sc NLLfast}.
For obtaining the $r$ values we used the numbers provided by {\sc CheckMATE},  including a
20\% uncertainty for the theoretical cross sections, which corresponds to the value used in \smodels\ when computing likelihoods.

In Table~\ref{tab:CMresults} we show, for this specific pMSSM point, the main contributions to the total signal yield for the 
best signal region (SR), $2jt$, in ATLAS-SUSY-2013-02. 
As we can see, if we only consider the symmetric $\tilde g \tilde g \to 2$~jets + $\MET$ topology,
we obtain an $r$ value very similar to the one obtained by \smodels\ ($r = 0.37$)
and the point is far from being excluded.
However, if we include the asymmetric $\tilde g \tilde q\to 3~{\rm jets} + \MET$ topology, the $r$ value increases
to $1.38$ and the point can be excluded. In contrast, the contribution from $\tilde q \tilde q$ production 
with $\tilde q\to q\tilde g$, $\tilde g\to g\tilde\chi^0_1$ has a tiny effect. 
For completeness, in the last line of Table~\ref{tab:CMresults} we
also present the inclusive result, which incorporates all possible gluino and squark decays, thus giving
a slightly higher $r$-value ($r=1.7$).

In this example we can clearly see that the asymmetric $\tilde g \tilde q$ topology is the dominant one and 
essential for excluding the tested point. 
This illustrates how SMS results for $\tilde g \tilde q$ topologies, with unrelated gluino and squark masses, 
would help improve the coverage of the pMSSM. 
Particularly useful would be efficiency maps, as they allow to combine different contributions to the same signal region 
in the simplified model context.

\begin{table}\centering
\begin{tabular}{lccc}
Topology & Cross section & Contribution to $2jt$ & $r$-value \\ \hline
$\tilde g \tilde g \to 2$~jets + $\MET$ & 250~fb & 21\%  & 0.37 \\
$\tilde g \tilde q \to 3$~jets + $\MET$ & 664~fb & 59\%  & 1.01  \\
$\tilde q \tilde q \to 4$~jets + $\MET$ & 136~fb & 4\%   & 0.08 \\ \hline
Total $\left(\tilde g \tilde g + \tilde g \tilde q + \tilde q \tilde q \right)$ & 1220~fb  & 100\% & 1.70 \\
\end{tabular}
\caption{Contributions of specific signal topologies to the total exclusion for the pMSSM point no.~192342466. 
The second column shows the topology cross section (production cross section times branching ratios),
while the third column shows the topology contribution to the signal yield for the $2jt$ signal
region of Ref.~\cite{Aad:2014wea}. The last column shows the $r$-values obtained using each topology individually.
The last line shows the corresponding results including all possible gluino and squarks decays, resulting in a
larger total cross section and $r$-value. See text for details. 
\label{tab:CMresults}}
\end{table}

\end{appendix}

\clearpage
\providecommand{\href}[2]{#2}\begingroup\raggedright\endgroup

\end{document}